\documentclass[journal]{IEEEtran}

\usepackage{cite}

\ifCLASSINFOpdf
   \usepackage[pdftex]{graphicx}
   \graphicspath{{../pdf/}{../jpeg/}}
   \DeclareGraphicsExtensions{.pdf,.jpeg,.png}
\else
\fi
\UseRawInputEncoding
\usepackage{amsmath}
\usepackage{array}
\usepackage{amssymb}
\usepackage{bbold}
\usepackage{standalone}
\usepackage{etoolbox}
\newtoggle{arXiv}%\toggletrue{arXiv}
\newtoggle{bibrun}%\toggletrue{bibrun}
\usepackage{graphicx}
\usepackage{color, calc}
\usepackage{array, booktabs}
\newcolumntype{L}{>{$}l<{$}}
\newcolumntype{C}{>{$}c<{$}}
\newcolumntype{R}{>{$}r<{$}}
\usepackage{tikz, tikz-3dplot}
\usetikzlibrary{arrows, arrows.meta, calc, positioning}
\usetikzlibrary{decorations.pathmorphing}
\tikzset{>=Stealth}
\usepackage{siunitx, physics}
\newcommand{\I}{\ensuremath{\mathrm{i}}}

\usepackage{enumitem}
\setlist[description]{labelindent=0pt, leftmargin=\parindent, font=\normalfont\itshape}
\usepackage{url}
\usepackage{subfigure}
\usepackage{textcomp}
\usepackage{eurosym}
\usepackage{xfrac}
\usepackage{pgfplots}
\usepackage{stmaryrd}
\pgfplotsset{compat=1.17}
\usepackage{makecell}
\usepackage{calrsfs}

\begin{document}
\title{Zero-echo-time sequences in highly inhomogeneous fields}

\author{\IEEEauthorblockN{
		Jose\,Borreguero\IEEEauthorrefmark{1}\IEEEauthorrefmark{2},
		Fernando\,Galve\IEEEauthorrefmark{1},
		Jos\'e\,M.\,Algar\'{\i}n\IEEEauthorrefmark{1}, and
		Joseba\,Alonso\IEEEauthorrefmark{1}}

		\IEEEauthorblockA{\IEEEauthorrefmark{1}MRILab, Institute for Molecular Imaging and Instrumentation (i3M), Spanish National Research Council (CSIC) and Universitat Polit\`ecnica de Val\`encia (UPV), 46022 Valencia, Spain}\\

		\IEEEauthorblockA{\IEEEauthorrefmark{2}Tesoro Imaging S.L., 46022 Valencia, Spain}\\

\thanks{Corresponding author: J. Alonso (joseba.alonso@i3m.upv.es).}}

% \markboth{Journal of \LaTeX\ Class Files,\,Vol.\,X, No.\,X, JANUARY\,2022}%
% {Shell \MakeLowercase{\textit{et al.}}: Bare Demo of IEEEtran.cls for IEEE Journals}

\maketitle

\begin{abstract}
	Zero-echo-time (ZTE) sequences have proven a powerful tool for Magnetic Resonance Imaging (MRI) of ultrashort $T_{2}$ tissues, but they fail to produce useful images in the presence of strong field inhomogeneities. Here we present a method to correct artifacts induced by strong $B_0$ inhomogeneities in non-Cartesian sequences, based on magnetic field maps obtained from the phase difference between two fast pointwise acquisitions. These are free of geometric distortions and can be used for model-based image reconstruction with iterative algebraic techniques. In this paper, we show that distortions and artifacts coming from inhomogeneites in $B_0$ are largely reverted by our method, as opposed to standard Conjugate Phase reconstructions. We base this technique on a ``Single-Point Double-Shot'' (SPDS) approach, and we have shown it to work even for intra-voxel bandwidths (determined by $B_0$ inhomogeneities) comparable to the encoding bandwidth (determined by the gradient fields). We benchmark the performance of SPDS for ZTE acquisitions, which can be exploited for e.g. dental imaging in affordable low-field MRI systems. Furthermore, SPDS can be used for arbitrary pulse sequences and it may prove useful for extreme magnet geometries, as in e.g. single-sided MRI.
\end{abstract}

%******************************************************************
%******************************************************************

 \ifCLASSOPTIONpeerreview
 \begin{center} \bfseries EDICS Category: 3-BBND \end{center}
 \fi
\IEEEpeerreviewmaketitle

%******************************************************************
%******************************************************************

\section{Introduction}\label{sec:Intro}

Producing diagnostically valuable images of  biological tissues with ultra-short (sub-millisecond) $T_{2}$ is a long standing goal in MRI research \cite{Weiger2019}. Special-purpose pulse sequences such as UTE (Ultra-short Echo Time), ZTE (Zero Echo Time) and SWIFT (Sweep Imaging with Fourier Transformation) have been crucial, and can be employed for a variety of applications, including dental, lung, musculoskeletal or myelin imaging \cite{Idiyatullin2011,GEIGER2021708.e9,Chang2015,Hyungseok2020}. Among these sequences, ZTE imposes the most stringent constraints on hardware specifications, but it is best suited for MRI of tissues with sub-ms $T_{2}$ \cite{Weiger2012}.

The advent of clinically viable low-field MRI (LF-MRI) technologies has brought along a plethora of new possibilities which can be realized by translating clinical high-field techniques to the low field regime, at a lower cost and in a more accessible fashion \cite{Marques2019,Sarracanie2020,GuallartNaval2022}. Imaging of ultra-short $T_{2}$ tissues, however, is not one of them. While \emph{in vivo} dental human MRI has been demonstrated at $B_0\geq3$\,T \cite{Stumpf2020,Tymofiyeva2008}, it remains a pending challenge in the sub-tesla regime, mostly due to lack of SNR (Signal-to-Noise Ratio). ZTE-like sequences such as PETRA (Pointwise Encoding Time reduction with Radial Acquisition, \cite{GrodzkiPETRA}) have been used for simultaneous soft and hard tissue imaging at low fields (260\,mT), but only with \emph{ex vivo} samples \cite{Algarin2020}. Furthermore, these systems feature small FoVs (Field of View) and are incompatible with \emph{in vivo} clinical operation \cite{Gonzalez2021,Borreguero2023}. To explore the viability of hard-tissue {\emph{in vivo}} LF-MRI, we built a custom 197\,mT yoked magnet (see Appendix A) for which, due to geometric and weight constraints, the principal magnetic field is extremely inhomogeneous and standard reconstructions suffer from severe artifacts.

In this paper we first describe a method to measure field inhomogeneity in a reliable way, and show that the usual double-shot gradient-echo (GRE) technique \cite{Geiger2020} fails beyond usability due to irrecoverable intra-voxel phase accumulation and geometric misassignment of spatial coordinates. Our method, although more time-costly in certain contexts, can cope with extreme field inhomogeneities, leveraging a double-shot scheme with single-point acquisitions which accumulate a single phase per shot (Single-Point Double-Shot, SPDS). Using the acquired image-based $B_0$ map as prior knowledge (PK), we build an encoding matrix (EM) for algebraic reconstruction through Kaczmarz's algorithm (ART, \cite{Kaczmarz1937}) which yields faithful images in situations where Conjugate-Phase reconstruction \cite{Noll1991} fails. The combination of SPDS and ART works even near the limit of irrecoverability, where the applied encoding gradient spans an inter-voxel bandwidth (BW) similar to the intra-voxel bandwidth caused by $B_0$. We demonstrate the benefits of the method for linear and quadratic inhomogeneities, as well as for fields with finer spatial details and abrupt variations.

\section{Methods}\label{sec:Methods}

\subsection{MRI scanners and samples}\label{sec:Dental2}

\begin{figure*}
	\centering
	\includegraphics[width=\textwidth]{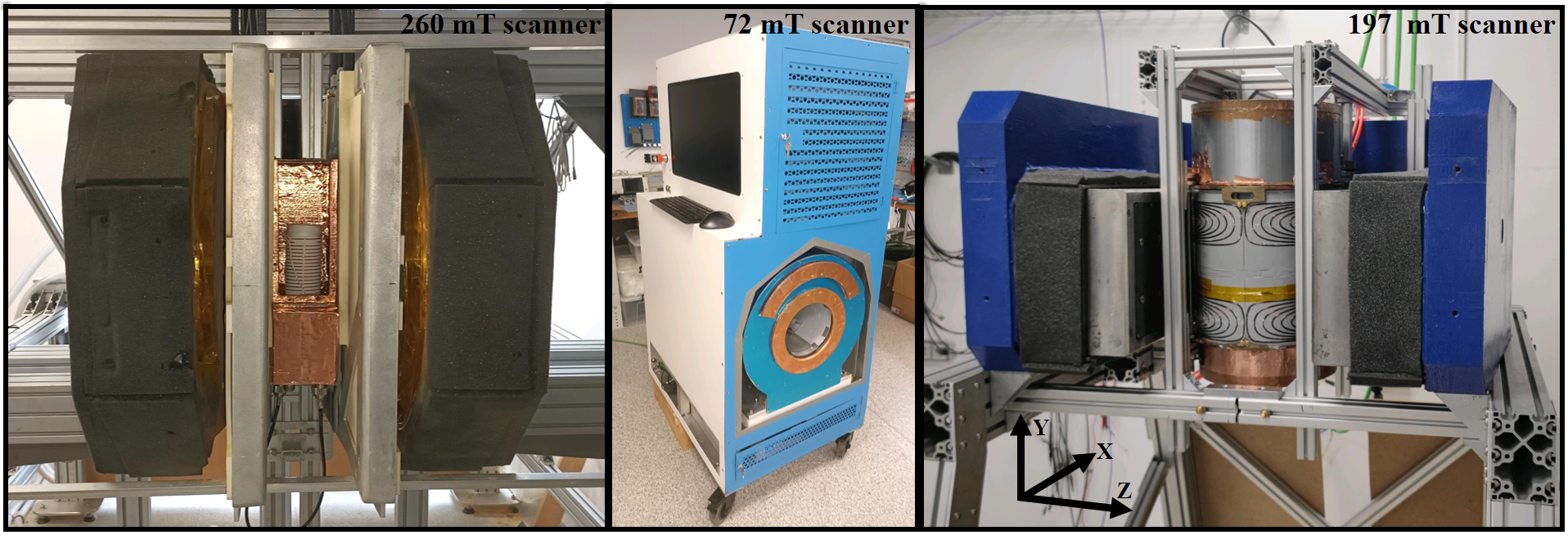}
	\caption{The three scanners used for our experiments. Left) 260\,mT scanner with 20\,ppm inhomogeneity in a 150\,mm diameter FoV. Middle) 72\,mT scanner with 3,100\,ppm in a 200\,mm diameter FoV. Right) 197\,mT scanner with 14,000\,ppm in a $92\times71\times110$\,mm$^3$ FoV.}
	\label{fig:fig_Scanners}
\end{figure*}

We have carried out experiments in the three MRI scanners shown in Fig.\,\ref{fig:fig_Scanners}): i) a highly homogeneous 260 mT C-shaped yoked magnet (described previously in \cite{Algarin2020}); ii) a reasonably homogeneous 72 mT Halbach array-based scanner (described previously in \cite{GuallartNaval2022}); and iii) a highly inhomogeneous 197\,mT scanner introduced below and described extensively in Appendix~A.

The latter scanner has a C-shaped yoked magnet providing a 197\,mT main field, where the gap between poles is 27\,cm and the total weight is around 1,200\,kg. The field inhomogeneity has been measured to be around 14,000\,ppm (parts per million) in a FoV of $92\times71\times110$\,mm$^3$, and the spatial distribution is mostly quadrupolar at its center (see Appendix\,\ref{sec:app}, Fig.~\ref{fig:Dental2scanner}). The encoding gradient coils are formed by wires wound around water-cooled copper plates, and can achieve a maximum of $\approx76, 27, 51$\,mT/m strengths (at 100\,A) along the $x, y, z$ directions, respectively. The transmission (Tx) coil of the radio-frequency (RF) system is a solenoid with $Q\approx52$ and can produce a $\pi/2$ pulse in $\approx\SI{16}{\micro s}$. We employ a variety of reception (Rx) coils, which we typically adjust for maximum SNR with different samples.

All phantoms used for reconstruction studies are replicas of a PLA 3D-printed hollow cylinder with the characters `i3M' inside, with 4\,cm outer diameter and character thickness 2-3\,mm. The exception is a $8\times13$\,cm$^2$ rectangular phantom with 8\,mm thick characters in Fig.\,\ref{fig:fig_SPDSexamples}f, due to the larger available FoV in the 72\,mT scanner. For $B_0$-mapping studies (Fig.\,\ref{fig:fig_SPDSvsClasicApproach}) we use an empty hexagonal phantom with 4\,cm outer diameter. Although we employ ZTE sequences, specifically PETRA, because we seek compatibility with ultra-short $T_2$ imaging, we fill the phantoms with a 1\,\% CuSO$_{4}$ water solution. This features $T_2\approx7.5$\,ms (measured with CPMG, \cite{Carr1954,Meiboom1958}) and $T_1\approx16$\,ms  (Inversion-Recovery, \cite{Bydder1998}) in the 260\,mT scanner. At any rate, for the ZTE sequences used in this article $T_2^{*}$ is most relevant and is in the range \SI{200}{\micro s} - 1\,ms for the cases considered here.

\subsection{Field estimation}\label{sec:FieldEstimation}

\begin{figure*}
	\centering
	\includegraphics[width=1\textwidth]{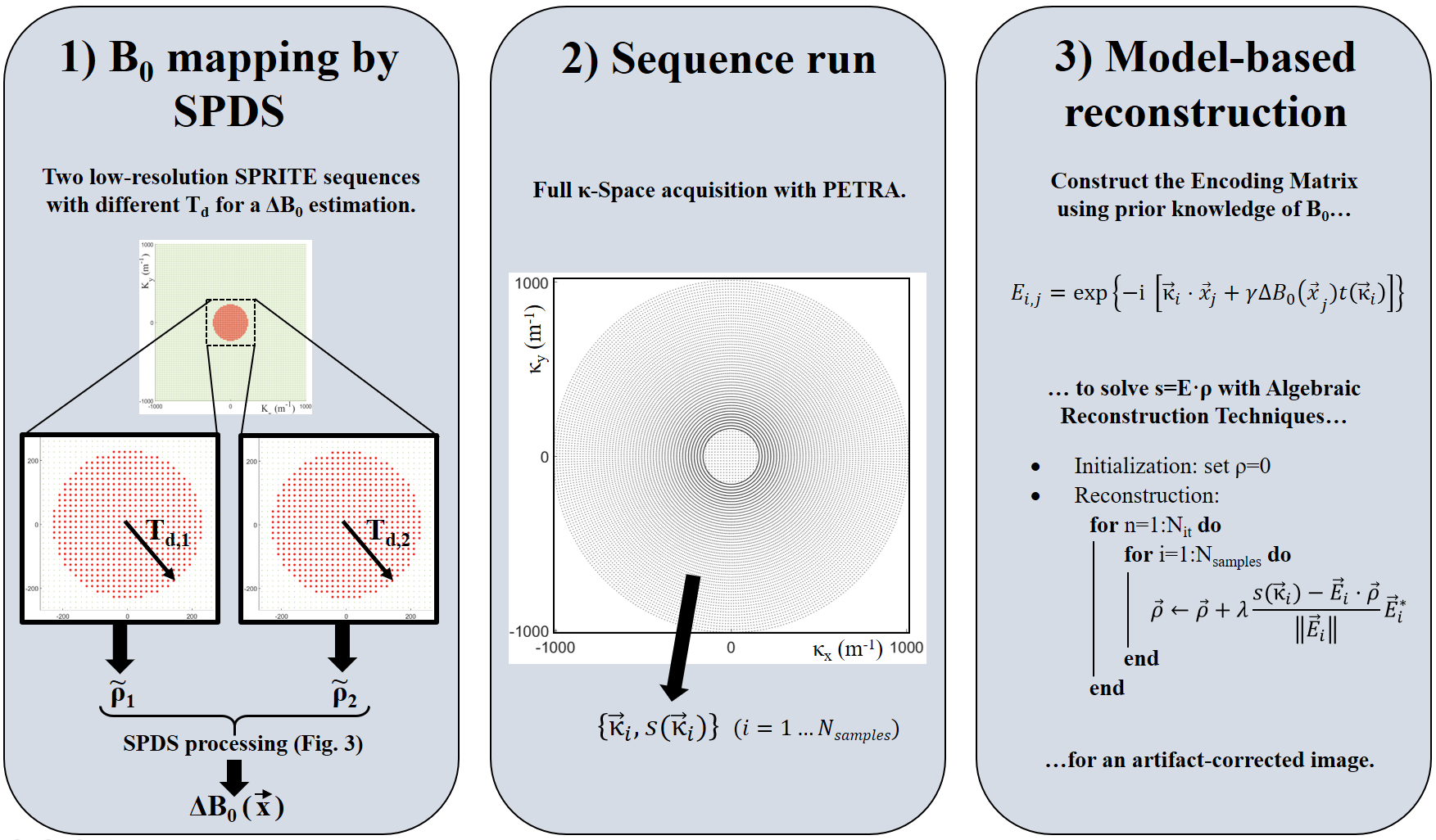}
	\caption{Diagram of the proposed procedure for ZTE imaging in highly inhomogeneous fields. 1) Field map estimation with by SPDS. 2) Full PETRA acquisition. 3) Model based reconstruction with the $\Delta B_0$ obtained in step 1).}
	\label{fig:fig_scheme}
\end{figure*}

Our model-based reconstruction leverages accurate knowledge of the $B_0$ field distribution. We estimate this with three different techniques: i) a Hall probe scanning the FoV volume guided by a robot; ii) the usual double-shot echo-time-shifted GRE \cite{Geiger2020}; and iii) our double-shot single-point Cartesian acquisition (SPDS), where each shot provides a different encoding time.

\subsubsection{Hall sensor}\label{sec:Hall}
For the first procedure we use a THM1176 \cite{Metrolab} magnetometer mounted on a home-made $xyz$ linear stage to measure $B_0(\vec{x})$ on a Cartesian grid over a volume 15$\times$10$\times$8\,cm$^{3}$ with isotropic spacing of 1\,cm. We strived to reference the $B_0$ map to image space by finding the saddle point of the measured field and matching it to the image, even if this turned out to be experimentally challenging.

\subsubsection{Double-shot GRE}\label{sec:GRE}
The double-shot GRE method uses two low-resolution images from two GRE sequences with one echo per repetition, where the echo time (TE) is different for each acquisition. Typically, the $k$-spaces are Fourier-transformed to produce images $\rho_{1}$ and $\rho_{2}$. Since most of the signal energy is concentrated on the echo, one usually assumes that the phases (arg) of the images differ by $\approx\gamma \Delta B_{0} (\text{TE}_2-\text{TE}_1)$ in all voxels, where $\gamma$ is the $g$-factor and $\Delta B_{0}$ is the off-resonance at each voxel. The field can then be estimated as:
\begin{equation}
\Delta B_{0}\approx\frac{\text{arg}(\rho_{2})-\text{arg}(\rho_{1})}{\gamma \left( \text{TE}_\text{2}-\text{TE}_\text{1} \right)}.
\label{eqn:B0fromGE}
\end{equation}

GRE sequences are inherently sensitive to field inhomogeneity. To minimize geometric distortions, it is convenient to fix TE and the readout window duration as short as possible. However, this is ultimately limited by hardware, hence images invariably suffer from non-negligible geometric distortions in our regime of extreme off-resonance, leading to incorrect assignments of spatial coordinates.

\subsubsection{SPDS}\label{sec:SPDS}
Our SPDS procedure also requires two acquisitions for field mapping, but it does not rely on echoes. Instead, each shot is a low-resolution Cartesian acquisition (\`a la SPRITE, \cite{BALCOM1996131}) where every point is acquired with the same encoding time (i.e. $T_{\text{d,}i}$\footnote{We import the term ``dead time'' $T_\text{d}$ from ZTE for SPDS, as it is also determined by electronic ring-down and blanking effects.}, see panel 1 in Fig.~\ref{fig:fig_scheme}). These can be reconstructed either with ART or with an FFT. All $\kappa$-space\footnote{In this work we intentionally use the notation $\kappa$-space, rather than $k$-space, to highlight the fact that off-resonant effects are extreme and preclude most well-known $k$-space properties, such as the fact that resolution is constant across an image and is determined by $k_\text{max}$. This is further discussed in Appendix~\ref{sec:app3}.} points in a given acquisition are subject to the same phase accumulation due to inhomogeneities. In this sense, each image has a unique phase in every voxel, and thus the phase difference between the two images is an exact proxy of the $B_0$ field. Note that the signal equation for a single-shot is:
\begin{equation}
s(\vec{\kappa})=\int \text{d}\vec{x}\, \text{e}^{-\I\vec{\kappa}\cdot\vec{x}} \text{e}^{-\I\gamma\Delta B_0(\vec{x})t(\vec{\kappa})}\rho(\vec{x}),
\end{equation}
with $\Delta B_0(\vec{x})$ the deviation of the local field at position $\vec{x}$ with respect to $B_0$ (i.e. the inhomogeneity), and $t(\vec{\kappa})$ the time between spin excitation and the acquisition of signal at point $\vec{\kappa}$. Since all $t(\vec{\kappa})$ are equal for each acquisition, that is equivalent to $\rho_i$ (the low-resolution ART reconstruction corresponding to shot $i$) having an extra global phase $-\I\Delta B_0(\vec{x})T_{\text{d,}i}$, i.e.
\begin{equation}
s_i(\vec{\kappa})=\int \text{d}\vec{x}\, \text{e}^{-\I\vec{\kappa}\cdot\vec{x}}\tilde{\rho}_i(\vec{x}),
\end{equation}
with
\begin{equation}
\tilde{\rho}_i(\vec{x})=\text{e}^{-\I\gamma\Delta B_0(\vec{x})T_{\text{d,}i}}\rho(\vec{x}).
\end{equation}
The specific choice of a {\it{global}} $T_{\text{d,}i}$ for all signal acquisition points $\vec{\kappa}$ makes SPDS immune to geometric distortions due to $B_0$, and the field map can be obtained as
\begin{equation}
\Delta B_{0}=\frac{\text{arg}(\tilde{\rho}_{2})-\text{arg}(\tilde{\rho}_{1})}{\gamma \left( T_\text{d,2}-T_\text{d,1} \right)}.
\label{eqn:B0fromSPDS}
\end{equation}

Notably, if these acquisitions are Cartesian and fully sampled, a discrete Fourier transform can be used. Furthermore, this method is compatible with arbitrary non-Cartesian sequences, as long as $t(\vec{\kappa})$ is the same in both shots. Key to this discussion is that Eq.\,(\ref{eqn:B0fromSPDS}) is an {\it{exact}} equality, while Eq.\,(\ref{eqn:B0fromGE}) is an approximation which degrades for stronger inhomogeneities.

After completion of this work, we found a sequence similar to SPDS published in a PhD thesis (mSPRITE, \cite{bazzi2022}). The authors use mSPRITE to map $\Delta B_{0}$ distortions (up to 150\,ppm) generated by metallic implants in an otherwise highly homogeneous $B_0$. To this end, they take a set of five $\kappa$-space points by measuring at different $T_{\text{d,}i}$ within each readout gradient pulse, and produce five fully-sampled SPRITE images. The off-resonance map can then be obtained by fitting the phase evolution for each image pixel as a function of $T_{\text{d,}i}$. This is a single-shot approach and therefore faster than SPDS. However, their $\kappa$-space sampling and spacing differ among acquisitions, precluding simple Fourier analysis, affecting the resolution and FoV of the final field map, and thereby compromising its fidelity and the overall method performance.

For both GRE and SPDS, we use a masking procedure where we only take $B_0$-map points for which $|\rho|$ is above a certain threshold. In this way we remove background noise. In our experiments these threshold values were adjusted individually for optimal results, but this can be automatized. After masking, each image is phase-unwrapped with a MATLAB plugin \cite{unwrapMullen}. Finally, the pixel-based $B_0$ map is fitted to a 2$^\text{nd}$ to 5$^\text{th}$ order polynomial, depending on the spread of the 95\,\% confidence bounds of the coefficients and its performance when the polynomial function is used for ART-based reconstruction. Note that low-order fits lead to reduced Gibbs ringing in the resulting field maps.

The number of acquired $\kappa$-space points ($\kappa_\text{SPDS}^\text{max}$) restricts the resolution of the final map, and has to be adjusted depending on the smoothness of the magnetic field inhomogeneities to be mapped, as well as on the length scale of relevant structures in the sample. We have found that 600 to 1,200 points, reconstructed in a matrix size of 120 $\times$ 120 and FoV 6$\times$6\,cm$^{2}$ (resolutions in the range between 2.4\,mm and 1.6\,mm), suffice to resolve $\Delta B_{0}$ and faithfully mask the internal letters of the phantoms employed in this paper (2\,mm thickness).

The procedure for field estimation by SPDS is depicted in Fig.~\ref{fig:fig_scheme} (panel 1). The first SPDS maps were taken in the 260\,mT scanner after intentionally spoiling the field by means of a single N48 grade NdFeB cubic magnet ($B_\text{rem}\approx1.4$\,T) of size (4\,mm)$^3$, placed $\approx1.5$\,cm away from the sample. This generates a field variation of $\approx\SI{250}{\micro T}$ within the circular phantom of diameter $\approx40$\,mm. Before image acquisition, SPRITE sequences were executed consecutively with different dead times, $T_\text{d,1}=\SI{175}{\micro s}$ and $T_\text{d,2}=\SI{250}{\micro s}$, acquiring a set of 788 samples measured in the same $\kappa$-space values and centered on a circumference of radius 250\,m$^{-1}$. These datasets are reconstructed using ART with 10 iterations and $\lambda=0.1$ in a FoV of (6\,cm)$^2$ with a matrix size of $120\times120$. After that, the phase of the complex images is determined and masked, removing pixels with intensities $<50$\,\% of the maximum absolute value of the image. Hereafter, we unwrap the phase and use Eq.~(\ref{eqn:B0fromSPDS}) to estimate $\Delta B_{0}$ in the selected points. Finally, the data $\lbrace \vec{x}_{i}, \Delta B_0(\vec{x}_{i}) \rbrace$ are fitted to a 5$^\text{th}$ degree polynomial. The time needed to reconstruct both datasets, phase unwrapping, masking, phase subtraction and fitting is $<5$\,s in a standard CPU (Intel(R) Core(TM) i7-6700 CPU @ 3.40\,GHz). Note, however, that this depends strongly on the number of sampled points and can hence change for different FoV, phantoms structures, 3D acquisitions, etc. SPRITE acquisitions take $\sim40$\,s each.

\subsubsection{GRE vs SPDS}\label{sec:GREvsSPDS}
We also compared the field estimations by GRE and SPDS for varying degrees of inhomogeneity. Each scenario of inhomogeneity is generated by changing the configuration of N48-grade NdFeB blocks in the  260\,mT scanner, either adding more or simply bringing the magnets closer to the sample. To assess $\Delta B_0$ in each situation, GRE images of the hexagonal phantom (see Sec.\,\ref{sec:Dental2}) have been acquired in a FOV = (6\,cm)$^2$ in a matrix size of $N=60\times60$, with full $\kappa$-space coverage and Fourier reconstruction. All GRE images have $T_\text{acq}=0.5$\,ms and the lowest TE of each pair has TE$_1=1.55$\,ms, approaching the maximum allowed by gradient coils and amplifier; meanwhile, TE$_2$ is chosen seeking a compromise between phase-accumulation, which improves precision, with phase differences $<2\pi$ between shots, which would otherwise invalidate Eq.\,(\ref{eqn:B0fromSPDS}). We average longer for increasing inhomogeneities to compensate for SNR loss due to $T_{2}^{*}$ shortening: $N_\text{avg}=50$, 250 and 400 (scan times of 2.5, 12.5 and 20\,min). SPRITE images have been acquired in a FOV = (6\,cm)$^2$ with a partial, low-frequency coverage of $\kappa$-space, reconstructed into a matrix size of $N=60\times60$ with ART. Similarly to the GRE case, in SPDS we choose $T_\text{d,1}$ and $T_\text{d,2}$ and the number of acquired points according to the phase accumulation in each case, measuring $N_\text{SP}=235$, 521 and 492 points inside a circumference of radius 176, 214 and 210\,m$^{-1}$. Here we also average depending on the inhomogeneity: $N_\text{avg}=9$, 16 and 50, for a total scan time of 2.8, 7 and 20\,min for each shot, respectively.

\subsection{Data acquisition and image reconstruction}\label{sec:Reconst}
\subsubsection{ART}
Once a field map has been estimated, we fit $\Delta B_0$ to a polynomial function and run a full $\kappa$-space PETRA acquisition, where the fitted function is used in the EM:
\begin{equation}
\label{eqn:encodingmatrix}
E_{i,j}=\exp \left\lbrace  -\I \left[   \vec{\kappa}_i\cdot\vec{x}_j +\gamma\Delta B_{0}(\vec{x}_j)  t(\vec{\kappa}_i)   \right]      \right\rbrace.
\end{equation}
The signal equation is thus $s=E\,\rho$. In our PETRA acquisitions, $t(\vec{\kappa})=T_\text{d}$ for the central point-wise region, and $t(\vec{\kappa})=|\vec{\kappa}|/(\gamma G)$ for the radial part. For benchmarking purposes, we often reconstruct images omitting the prior knowledge, i.e. we use the EM without the $\Delta B_{0}$ term. All reconstructions in this paper employ the Julia Programming Language \cite{Julia}, with \texttt{CUDA.jl} for GPU-acceleration in an Nvidia GeForce GTX2080Ti card, with iterations $N_\text{it}=10$ and update parameter $\lambda=0.1$ (which takes $\sim3$ seconds). The only exception is the 3D acquisition in Fig.~\ref{fig:fig_3DSPDS} where, due to the computational time required, we set $N_\text{it}=\lambda=1$ ($\sim10$ minutes).

\subsubsection{Simulations}
We start by testing the performance of ART with prior knowledge coming from SPDS in fields with known distribution. To this end, we simulate the effect of running PETRA sequences on a 2D Shepp-Logan digital phantom subject to field distributions akin to those we use later in experiments. To prevent reconstruction crimes\cite{crime2004}, the simulated signal takes the contribution of 100 spins per reconstruction pixel.

These simulations use Julia in the same GPU card as for reconstruction. Typical signal simulation times are $\sim15$\,min, and $\sim8$\,s for one iteration of ART reconstruction.

\subsubsection{2D projections}\label{sec:ReconstExp}
A goal of this work is to show model-based ART reconstructions for different inhomogeneity scenarios and scanners. To this end, all PETRA and SPRITE images have been acquired in 2D. We consider different cases of $\Delta B_0$ inhomogeneity: the 260\,mT scanner (20\,ppm inhomogeneity over a spherical region of 150\,mm in diameter); the 260\,mT scanner, where we have superimposed a linear 15\,mT/m gradient field along the $z$-axis, i.e. around half the value of the encoding gradient ($\approx29.4$\,mT/m); the 260\,mT scanner, where we place a NdFeB cube of N48 grade ($B_\text{rem}=1.4$\,T) with volume (4\,mm)$^3$, similarly to Fig.\,\ref{fig:fig_SPDSprocedure}; the 197\,mT scanner (roughly quadrupolar with curvatures $\approx(0.2,-0.7,0.5)$\,T/m$^{2}$), with the phantom along the $xz$ and $yz$ planes; and the 72\,mT scanner \cite{GuallartNaval2022}. For each situation, we run  a PETRA sequence and perform SPDS to obtain a $\Delta B_0$ estimation. Except for the 72\,mT system, PETRA datasets were acquired in a FOV of 6$\times$6\,cm$^2$ with full $\kappa$-space radial coverage according to the Nyquist criterion and reconstructed into a matrix of 120$\times$120 with ART. For these, we set $T_\text{d}=\SI{100}{\micro s}$ and $T_\text{acq}=\SI{800}{\micro s}$ for a maximum readout gradient strength of $\approx29.4$\,mT/m. The PETRA dataset in the 72\,mT scanner was acquired in a FOV of (20\,cm)$^2$ with full $\kappa$-space radial coverage reconstructed into a matrix of 150$\times$150 with ART. Due to longer coil ring-down times and stronger heat dissipation at the gradients, we set $T_\text{d}=\SI{380}{\micro s}$ and $T_\text{acq}=4$\,ms for a maximum readout gradient strength of $\approx2.2$\,mT/m. In all $\Delta B_0$ situations, we use a polynomial model to fit the pairs $\lbrace \vec{x}_{i}, \Delta B_0(\vec{x}_{i}) \rbrace$ coming from SPDS, considering different orders and checking their performance when used for ART reconstruction with prior knowledge.

\subsubsection{3D acquisition}\label{sec:3DExp}
We carried out also a model-based ART reconstruction for a 3D acquisition in the 260\,mT scanner. A first dataset was obtained after degrading $\Delta B_0$, again by close proximity of a N48 cube of NdFeB, and a second dataset was acquired in the homogeneous setting, after removing the NdFeB cube. Both datasets were fully sampled according to the Nyquist criterion, in a FoV of (6\,cm)$^3$ reconstructed into 120$^3$ voxels, $T_\text{d}=\SI{100}{\micro s}$, $T_\text{acq}=\SI{800}{\micro s}$, $G_\text{readout}\approx29.4$\,mT/m, $N_\text{radial}=45,332$, $N_\text{single}=2,872$ ($\kappa$-space samples for radial and single-point readouts, respectively), $\text{TR} = 50$\,ms, $N_\text{avg}=9$ and $T_\text{scan}\approx6$\,h. Before removing the N48 magnet, we generate a 3D map of $\Delta B_0$ with SPDS, where the SPRITE sequences had $T_\text{d,1}=\SI{150}{\micro s}$ and $T_\text{d,2}=\SI{200}{\micro s}$, measuring a total of 8,240 single points in a sphere of radius 210\,m$^{-1}$ after 5 averages, for an overall scanning time of $\approx34$\,min. The resulting SPDS raw data was fitted to a 5$^\text{th}$-order polynomial.

\subsubsection{CP vs ART}\label{sec:CPvsART}
We are not aware of previous comparisons between model-based reconstructions and Conjugate Phase reconstructions (CP, \cite{CP1985}, often used in non-Cartesian MRI) in the presence of strong field inhomogeneities. In this paper we strive to quantify their ultimate performances.

For this comparison, we used the 260\,mT scanner, with increasing linear inhomogeneity corresponding to intravoxel spectral bandwidths $\text{BW}_\text{vox}=0$, 106, 212, 426, 532, 639\,Hz/voxel (readout gradient strength of 0, 5, 10, 20, 25, 30\,mT/m). We generate these inhomogeneities with intentional DC offsets on the currents through the $z$ gradient coils. The PETRA acquisitions have common parameters: FoV = (6\,cm)$^2$, $N=120\times120$, $T_\text{d}=\SI{100}{\micro s}$, $T_\text{acq}=\SI{800}{\micro s}$, $\text{BW} = 75$\,kHz, $G_\text{readout}\approx29.4$\,mT/m, $N_\text{radial}=376$, $N_\text{single}=232$, $\text{TR} = 50$\,ms, $N_\text{avg}=10$ and $T_\text{scan}=5$\,min. The fixed 29.4\,mT/m encoding gradient is equivalent to a spectral separation of 625\,Hz/voxel. For completeness, we compare for each level of induced inhomogeneity (and therefore BW$_\text{vox}$) the reconstrucion of the corresponding dataset using prior-knowledge based ART against the well-known Conjugate Phase method \cite{CP1985}. To do so, we first interpolate the signal in $\kappa$-space (also $t(\vec{\kappa})$) into a rectangular grid with \texttt{ScatteredInterpolation.jl} in Julia (see Sec.~\ref{sec:barrierofthereconstruction}). After that, we follow the usual procedure, i.e. $$\rho(\vec{x}_j)=\sum_i (E_{i,j})^* s(\vec{\kappa}_i),$$ with the $\Delta B_0$ inhomogeneity term in the conjugate phase's exponent, that is using Eq.\,(\ref{eqn:encodingmatrix}) for $E_{i,j}$.

\section{Results}\label{sec:Results}

\subsection{Field estimation}\label{sec:SPDSexample}
\subsubsection{Hall sensor}
Mapping $\Delta B_0$ with a robot-mounted Hall probe, as described in Sec.~\ref{sec:Hall}, proved to be experimentally challenging. Together with the need for a free parameter in the EM to relate the RF demodulation frequency (which we typically set to the center of the FID spectrum) to the real Larmor frequency at the field center, all our attempts were unsatisfactory and only led to negative results.

\subsubsection{SPDS and GRE}
\begin{figure*}
	\centering
	\includegraphics[width=1\textwidth]{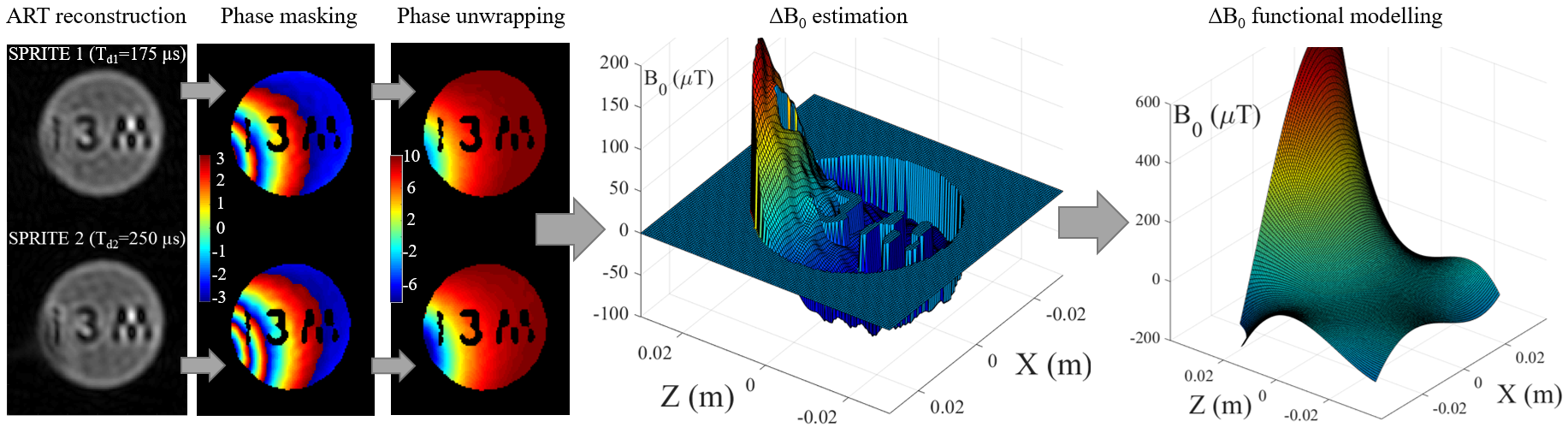}
	\caption{Steps for SPDS mapping of the field inhomogeneity $\Delta B_0$. The first step is the acquisition and reconstruction of low-resolution images with unique encoding times $T_\text{d,1}=\SI{175}{\micro s}$ and $T_\text{d,2}=\SI{250}{\micro s}$. Then, masking eliminates points from both images where signal intensity is below a given threshold with respect to maximum density of each image (50\,\% in this case). With the remaining points, we unwrap the phase and use Eq.~(\ref{eqn:B0fromSPDS}) to estimate $\Delta B_0$ at those points. Finally, we fit a 5$^\text{th}$ order polynomial to $\Delta B_0$.}
	\label{fig:fig_SPDSprocedure}
\end{figure*}

Figure~\ref{fig:fig_SPDSprocedure} shows the experimental results of running the SPDS procedure for field mapping in 2D (Sec.~\ref{sec:SPDS}).

\begin{figure*}
	\centering
	\includegraphics[width=1\textwidth]{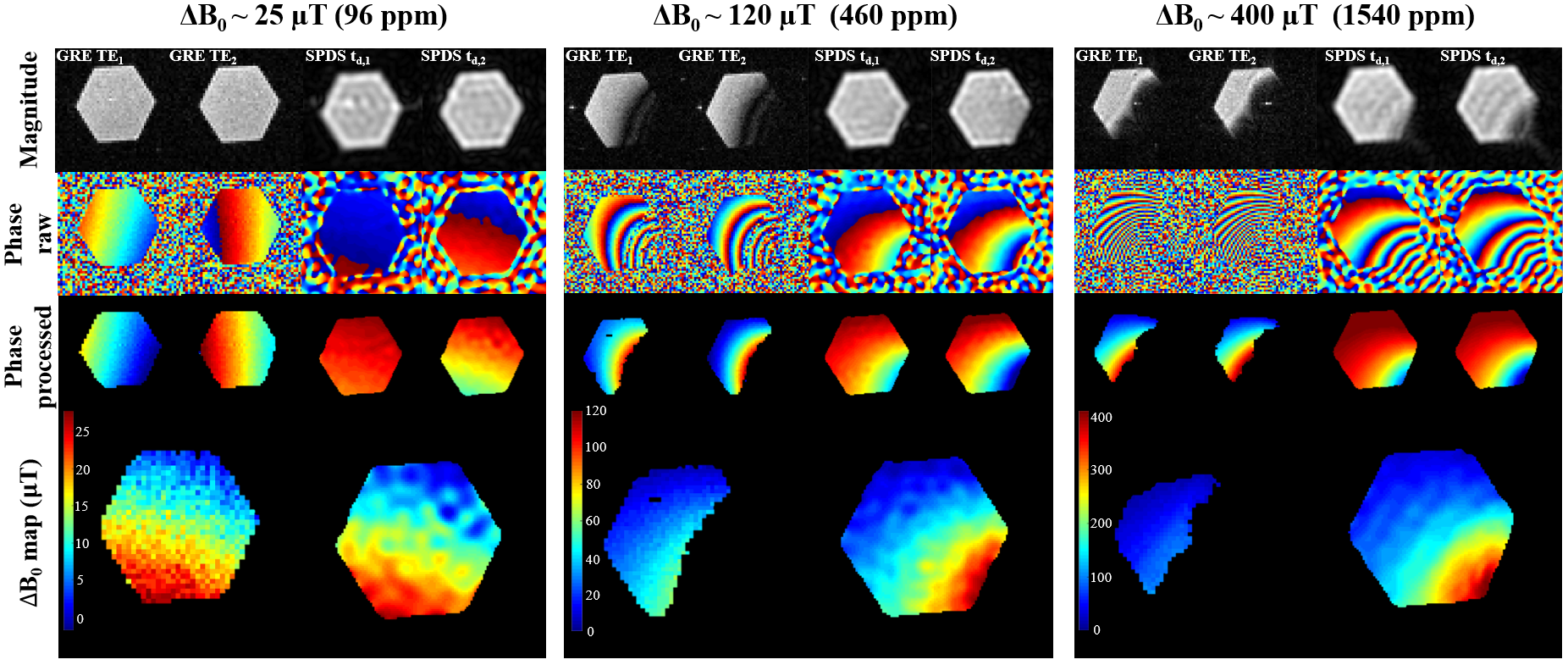}
	\caption{Comparison between SPDS vs double-shot GRE methods, for three different inhomogeneity levels and field arrangements. The upper row shows magnitude images corresponding to each shot. The second/third rows show phase images before/after masking and unwrapping. The lower row shows the estimated field inhomogeneity map.}
	\label{fig:fig_SPDSvsClasicApproach}
\end{figure*}

In Fig.~\ref{fig:fig_SPDSvsClasicApproach} we compare the performance of SPDS against the classic approach of double-shot GRE acquisition for different levels of induced inhomogeneity in the 260\,mT scanner (Secs.~\ref{sec:GRE} and \ref{sec:GREvsSPDS}). We plot the absolute value of the two-shot images, their wrapped and unwrapped phase and the estimated fields. Note that each column has a different configuration of extra NdFeB blocks that make $B_0$ intentionally inhomogeneous.

\subsection{Reconstruction with prior knowledge}\label{sec:SPDSresults}
\subsubsection{Simulations}
\begin{figure}
	\centering
	\includegraphics[width=0.5\textwidth]{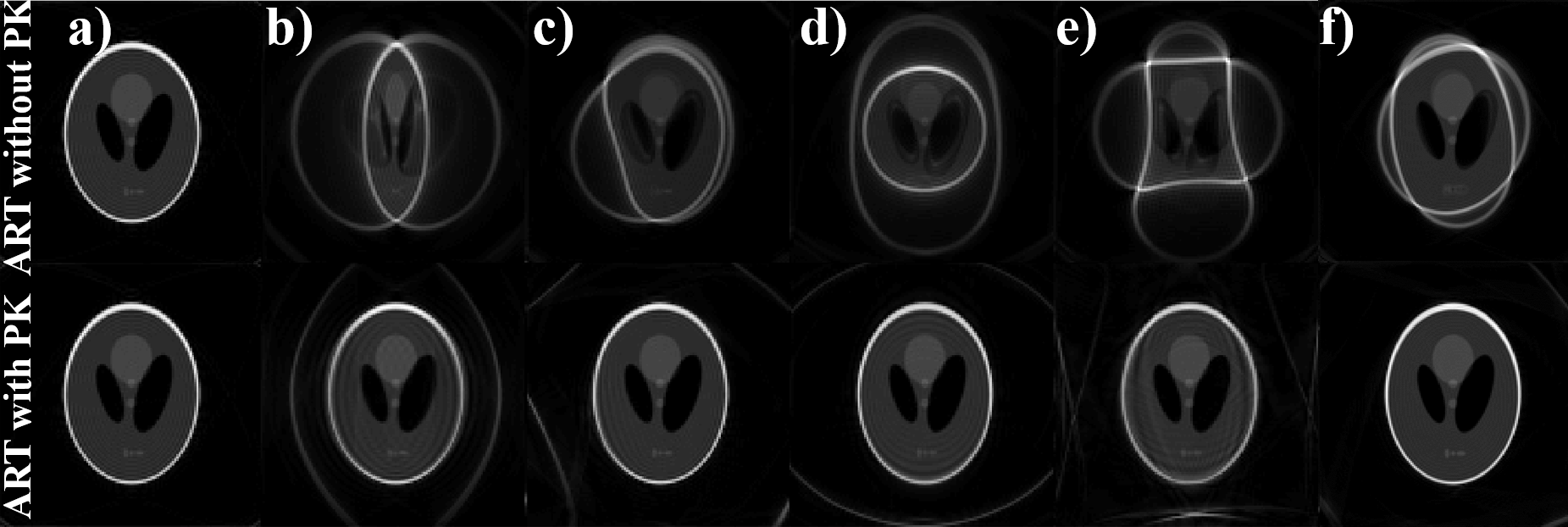}
	\caption{Performance of ART with perfect PK for field distributions explored experimentally, with a simulated Shepp-Logan signal with 100 spins per pixel.}
	\label{fig:fig_simulaciones}
\end{figure}

Figure~\ref{fig:fig_simulaciones} shows reconstructions of a simulated Shepp-Logan digital phantom with (bottom) and without (top) PK for field distributions similar to those used in the 2D experiments described in the next subsection.

\subsubsection{2D projections}
\begin{figure*}
	\centering
	\includegraphics[width=1\textwidth]{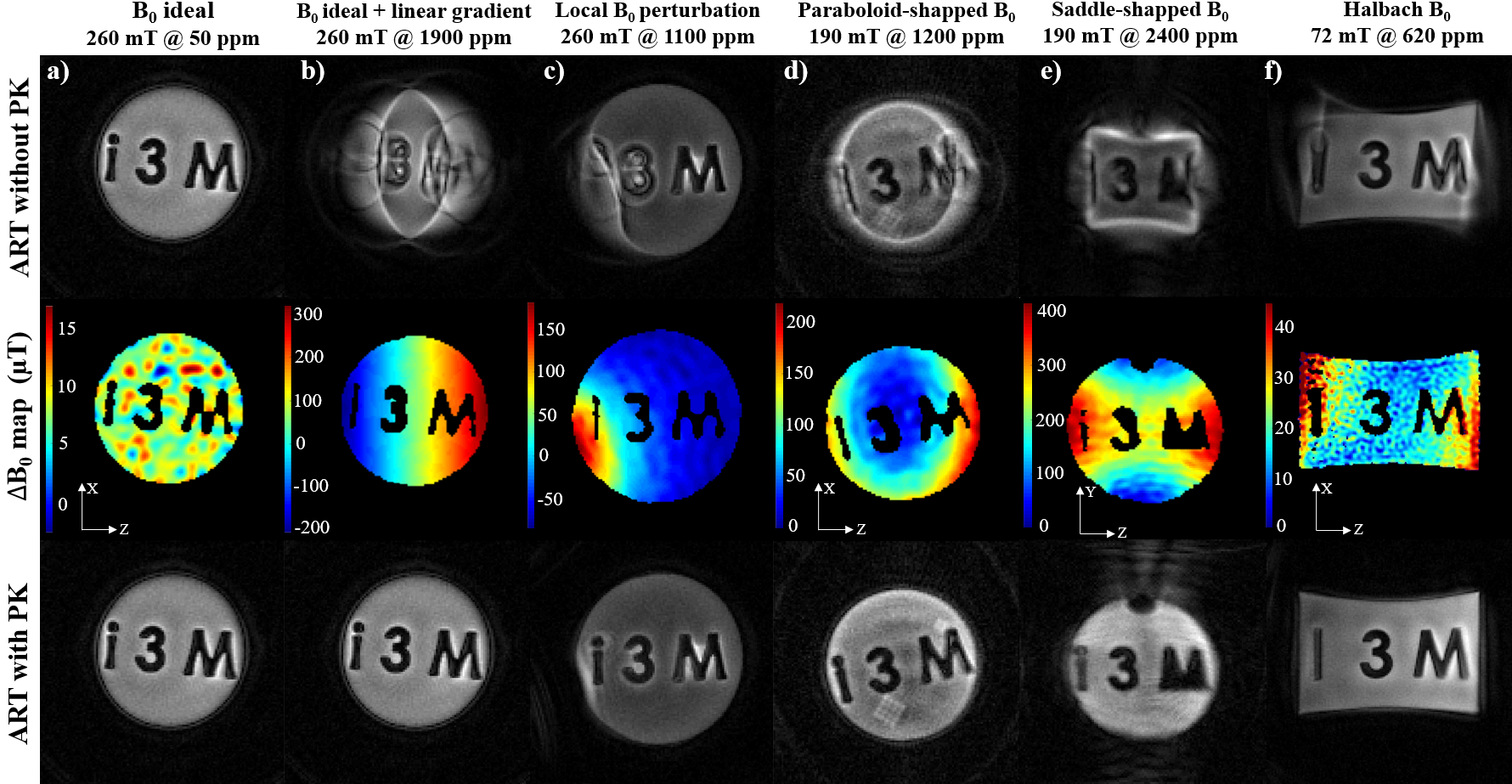}
	\caption{Performance of model-based ART reconstruction (bottom row) when prior knowledge of the field inhomogeneity comes from SPDS (middle row) for acquired 2D PETRA projections of a phantom in six different $\Delta B_0$ scenarios (see main text), as compared to reconstruction without PK (top row). There seems to be a small water infiltration inside phantom's PLA walls in d), leading to a grid-like structure. Also, in e) we can see the inlet for liquids in the phantom, since it is placed in a vertical position.}
	\label{fig:fig_SPDSexamples}
\end{figure*}

Figure \ref{fig:fig_SPDSexamples} shows that strong artifacts can be corrected by algebraic reconstruction when the SPDS field estimate is used as prior knowledge, in all three scanners and in situations where the field inhomogeneity is intentionally worsened by means of gradient fields or by bringing magnetic material close to the sample (see Sec.~\ref{sec:ReconstExp}).

For each $\Delta B_0$ scenario, we show ART reconstructions without prior knowledge (top row), the resulting $\Delta B_0$ map from an SPDS acquisition (middle row) and ART reconstruction including prior knowledge (bottom row). We ended up employing a 4$^\text{th}$-order polynomial for a) and c), 3$^\text{rd}$ order for f) and 2$^\text{nd}$ order for the rest, based on the perceived quality of reconstruction. A possible extension would be to use orthogonal polynomials at fixed expansion order, or direct interpolation of the measured map.

\subsubsection{3D acquisition}
\begin{figure}
	\centering
	\includegraphics[width=0.5\textwidth]{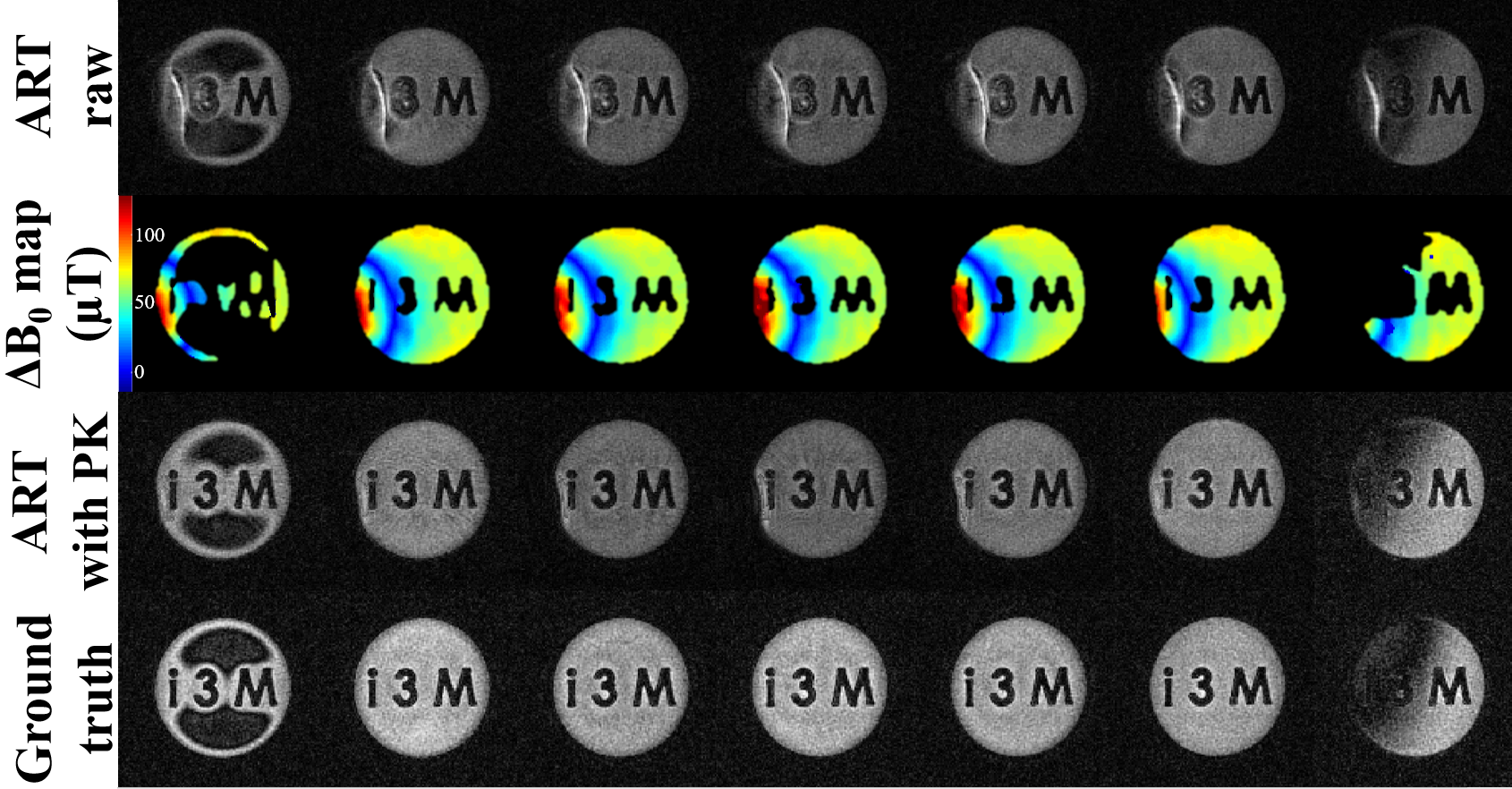}
	\caption{Perfomance of SPDS for 3D PETRA in the 260\,mT scanner with additional N48-grade NdFeB magnet cubes to force a non-linear field variation. In each column we show different $y$-slices from the top of the phantom (where capilarity effects are notorious) to the phantom base. We show (first row) dataset 1 reconstructed with ART but without PK of $\Delta B_0$, (second row) $\Delta B_0$ estimation in each slice obtained with 3D SPDS mapping, (third row) dataset 1 reconstructed with ART adding PK of $\Delta B_0$ coming from a 5$^\text{th}$-order polynomial fitting of a SPDS map shown previously. After removing magnet cubes, we show (last row) dataset 2 reconstructed with ART without prior knowledge of $\Delta B_0$, which represents the ground truth due to the homogeneity of $B_0$.}
	\label{fig:fig_3DSPDS}
\end{figure}

In Fig.~\ref{fig:fig_3DSPDS} we demonstrate the capabilities of SPDS mapping to create the prior knowledge of $\Delta B_0$ for a 3D PETRA acquisition (Sec.~\ref{sec:3DExp}) with different $y$-slices in each column of a phantom filled with CuSO$_4$ in the 260\,mT scanner. The ART reconstructions of dataset 1 with $\Delta B_0$ incorporated into the EM (third row) correct the large artifacts present in raw ART reconstructions without PK (first row) as compared to the ground truth (last row, where dataset 2 is reconstructed with ART without PK). A minor distorsion is seen towards the left region, where homogeneity is most compromised.

\subsubsection{CP vs ART}\label{sec:barrierofthereconstruction}

\begin{figure}
	\centering
	\includegraphics[width=0.5\textwidth]{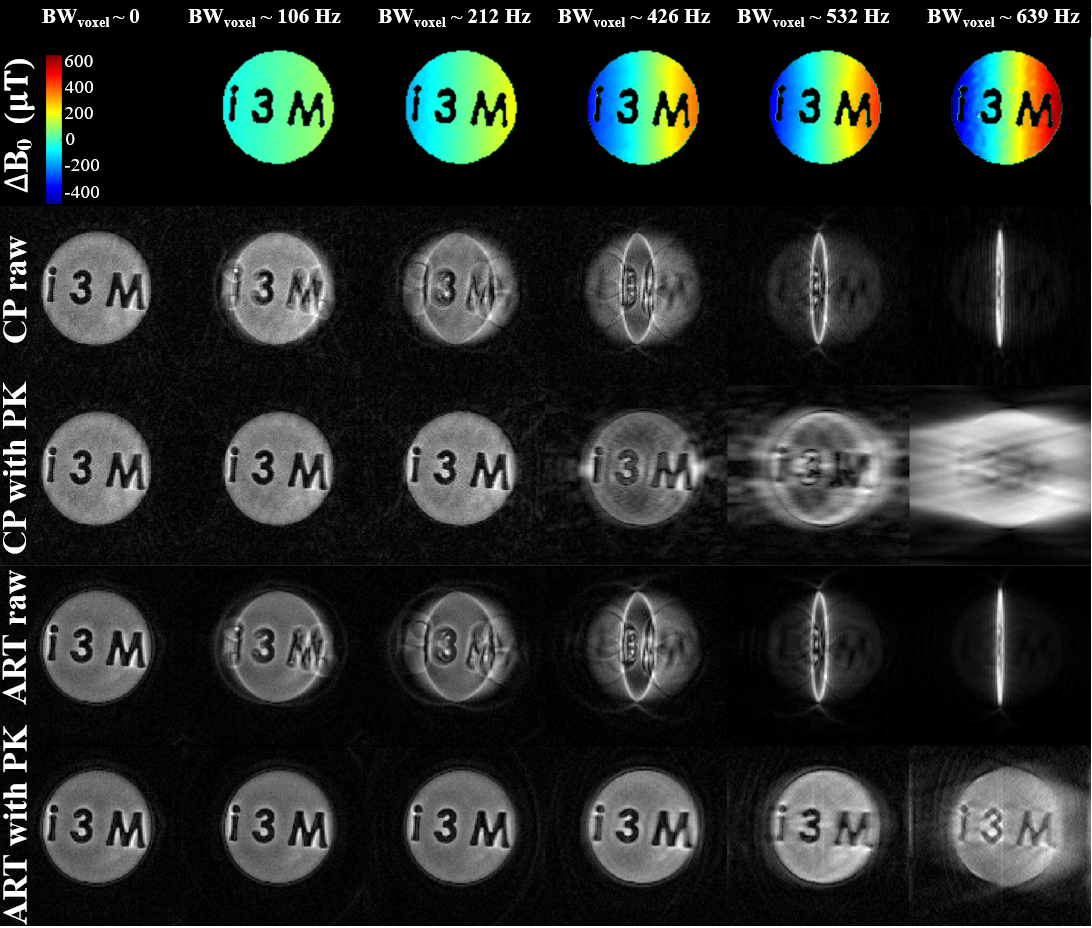}
	\caption{Experimental performance of ART and CP with increasing linear inhomogeneity, where SPDS (field fitted to 2$^\text{nd}$ order) is used to reconstruct PETRA datasets. All the images have been acquired with maximum gradient readout of $\approx29.4$\,mT/m, for an intravoxel bandwidth of 625\,Hz, while $\Delta B_0$ creates a frequency spread ranging from 0 to 639\,Hz. First row: $\Delta B_0$ mapping obtained with SPDS. Second row: CP without PK reconstructions. Third row: CP provided of PK reconstructions. Fourth row: ART without PK reconstructions. Fifth row: ART with PK reconstructions.}
	\label{fig:fig_ARTvsCP}
\end{figure}

Figure~\ref{fig:fig_ARTvsCP} shows CP and ART reconstructions in the presence of linear inhomogeneity up to and beyond the strength of the ZTE encoding gradient (Sec.~\ref{sec:CPvsART}). All reconstructions have been performed with the SPDS estimate as PK for the EM and, since all linear inhomogeneities are well known, also with the exact analytical form $\Delta B_0(z)= g_\text{inh.}\cdot z$, finding basically no difference between them. Thus, Fig.~\ref{fig:fig_ARTvsCP} is immune to imperfection in the field estimate, but the results shown include in the EM the field map obtained with SPDS.

\section{Discussion}\label{sec:discussion}

\subsection{Field estimation}\label{sec:disc_field}
The accuracy of field maps obtained with the THM1176 magnetometer is limited due to unknown tilts and oscillations of the rather long robotic arm, which is imposed by the scanner setup (see Fig.~\ref{fig:Dental2scanner}). We tried to suppress these errors by adjusting the type of trajectory used to fill the volume, the delay between probe positioning and field measurement, and the velocity of the motor in each step. However, the mechanical reference between the probe and image space coordinates (determined by the gradient coils) is not precise enough for image reconstruction, unless the probe is attached to some element which can be visualized with MRI. 

Additionally: i) if the encoding gradient fields are also inhomogeneous, physical coordinates and image-space coordinates (given by gradients) will not coincide far away from gradient centers, leading to further uncertainties; ii) the $\Delta B_0$ map is temperature dependent; iii) the map has to be obtained with sufficient resolution, which can require many hours; iv) due to the geometry of the scanner, the robotic arm length is significant and resulted in uncontrolled probe trajectories and oscillations during its displacement; and v) because the MRI signal is demodulated at an arbitrary frequency (typically the center frequency obtained by a simple free induction decay curve, FID), the Hall probe map must be shifted by a constant term which needs to be found. While in principle all these difficulties can be overcome, we found this method impractical. All these effects make MRI-based estimations of $B_0$ more reliable than direct field maps.

When comparing GRE with SPDS in Fig.~\ref{fig:fig_SPDSvsClasicApproach}, we observe that GRE is not reliable close to the source of high inhomogeneity, due to the large intravoxel $T_2^*$, and the image phase is not smooth in that region. Instead, SPDS works even for $\approx1,500$\,ppm (\SI{400}{\micro T} in 260\,mT). SPDS is limited by the time required to switch from trasmit to receive mode in the RF chain (dead-time), since $T_2^*$ leads to a drop in SNR in the meantime. On the other hand, GRE includes the additional duration of half the acquisition window, phase encoding and the gradient switching time. While SPDS requires the encoding gradient to be on before spin excitation (\`a la ZTE), GRE is based on gradient switching after RF excitation, and thus it is harder to shorten phase and encoding times due to inaccuracies given by e.g. eddy currents, leading to further hardware constraints. This highlights that, in situations of large inhomogeneities, the SNR will be a disadvantage for GRE. In addition, SPDS does not suffer from geometric distorsions in the field map, also favoring SPDS in medium-inhomogeneity regimes.

We also produced field maps with reduced $\kappa$-space sampling for all three configurations in Fig.~\ref{fig:fig_SPDSvsClasicApproach}. To this end, we acquired only $N_\text{SP}=120$ $\kappa$-space points (as compared to $N_\text{SP}= 235$, 521 and 492). This led to compatible results, except for $\Delta B_0\approx\SI{400}{\micro T}$, where the lower spatial resolution complicated masking around the bottom-right corner. The minimum number of $\kappa$-space samples is therefore determined by SNR level, masking and phase unwrapping stages, even if field inhomogeneities can be described by smooth (low-order) polynomials. The relevance of a reliable $\Delta B_0$ map is apparent from the comparison between Figs.~\ref{fig:fig_simulaciones} and \ref{fig:fig_SPDSexamples}, especially at borders or regions with abrupt field changes. A better map can be obtained either by increasing SPDS resolution (i.e. $\kappa_\text{max}^\text{SPDS}$), or by mapping a larger FoV with a bigger phantom. Another potential solution would be an iterative procedure, where the field estimate is updated at each step by consistency with the double-shot signals \cite{Fessler2004,Koolstra2021}. Still, one advantage of our method is that it does not require experimental iterations.

On the other hand, SPDS follows a pointwise scheme and is therefore inherently slower than GRE. Fortunately, $B_0$ distributions tend to be smooth and sampling small numbers of $\kappa$-space points usually suffices. This can lead to Gibbs ringing in borders, but low-order polynomial fits suppress such oscillations.

To minimize the acquisition time needed to perform SPDS, it is important to find a balance between the number of averages and the points sampled. The former helps to improve the SNR of the image, which is crucial to get reliable pairs $\lbrace \vec{x}_{i}, \Delta B_0(\vec{x}_{i}) \rbrace$, whereas the latter provides higher nominal resolution, facilitating the masking step in internal structured phantoms and enabling finer spatial details for cases where $\Delta B_0$ has abrupt variations. In Fig.~\ref{fig:fig_SPDSprocedure}, each SPRITE sequence lasts 39\,s for a total time of around 80\,s for the SPDS protocol. This could be reduced using shorter repetition times and Ernst angle excitation.

There is also a delicate balance when choosing the difference $T_\text{d,2}-T_\text{d,1}$. Ideally, this difference should be long in order to maximize the accuracy of $\Delta B_0$, but this can complicate the unwrapping procedure. Thus, depending on the unwrapping algorithm, more or less phase difference can be acquired for best field estimation, but the relative phase between shots cannot exceed $2\pi$, otherwise the obtained field would have an unknown scaling factor, as per Eq.\,(\ref{eqn:B0fromSPDS}). Furthermore, as can be seen in Fig.~\ref{fig:fig_SPDSprocedure} (right), the polynomial fits typically diverge outside of the fitted region. Thus, a calibration of the largest possible imaging region with a dedicated homogeneous phantom, with a sequence calibrated for maximum phase difference, is preferable.

In terms of total acquisition time for field estimation, the complete 3D SPDS protocol for Fig.~\ref{fig:fig_3DSPDS} took around 68\,min, plus several minutes for data treatment as previously explained (Fig.~\ref{fig:fig_SPDSprocedure}). While this may be excessively long for many applications, it is probably the only reasonable option in regimes of such extreme inhomogeneity. Note that this is still shorter than direct field mapping with a Hall probe, and a single calibration can be valid for a long time, as long as there are no significant field changes.

\subsection{Image reconstruction with PK}\label{sec:disc_reconst}

In all cases in Fig.~\ref{fig:fig_SPDSexamples} we see a major improvement as compared to reconstruction with no prior knowledge of the field, except for cases where the field changes abruptly such as c), where we see a small distorsion at the bottom-left border, or f) also near the borders (the image is non-rectangular due to non-linear gradients, but this could also be incorporated into the EM). The strongest artifacts are present in e), in addition to a small tilt of the sample which leads to a non-circular reconstruction nor SPDS. However, note that in b) the total field variation across the phantom is $\approx\SI{500}{\micro T}$ and reconstruction is close to ideal, while in e) a variation of $\approx\SI{400}{\micro T}$ introduces significant artifacts. Two possible reasons for this are: i) the SPDS estimate is not good enough; or ii) there is a hard physical limit for reconstructability even with perfect field knowledge. Simulations in Fig.~\ref{fig:fig_simulaciones} show no shape distorsion in the outer ellipse of the phantoms, and thus provide no explanation to the small deformation in the lower left part in Fig.~\ref{fig:fig_SPDSexamples}c). This means that SPDS is partially failing when the field has abrupt spatial changes. On the other hand, in Fig.~\ref{fig:fig_simulaciones}c) and e) we also observe artifacts within the phantom, precisely where the field changes most abruptly. We suspect that, at a given level of inhomogeneity, the encoding gradient is no longer able to spectrally resolve the difference between two voxels, and thus reconstruction is not possible even with perfect field knowledge.

\subsection{Limits of reconstruction}\label{sec:disc_limits}
Comparing the CP and ART reconstructions in Fig.~\ref{fig:fig_ARTvsCP} we observe that: i) conjugate phase fails sooner than ART, roughly when the inhomogeneity is half the value of the encoding gradient, while ART is able to reconstruct up to the level of the encoding gradient, even if it does so with severe artifacts; and ii) that despite perfect PK, massive artifacts are to be expected when the local inhomogeneity is at the encoding gradient strength.

\subsection{Cartesian and non-Cartesian acquisitions}\label{sec:further}
In Cartesian acquisitions \cite{NollReview2022} the time map can always be decomposed as $t(\vec{\kappa})=\text{const}+a\cdot \kappa_\text{readout}$, where the constant is related to the acquisition time before the given readout line is taken, and $a$ is a factor. In this case, the phase acquired due to inhomogeneity can be absorbed in a redefinition of the readout spatial coordinate $x'=x'(\vec{x})$, thus leading to geometric distortions. Hence, SPDS could be used to map $B_0$ accurately first, and then simply to map distorted coordinates $(x',y,z)$ into real coordinates $(x,y,z)$.
Contrarily, for non-Cartesian acquisitions, such coordinate mappings cannot be applied. In the case of radial acquisitions as in ZTE \cite{NollReview2022}, the point spread function (PSF) becomes circular with a radius proportional to $\Delta B_0/G$ for Fourier transform reconstructions. In Appendix\,\ref{sec:app2} we show the PSF for CP and ART reconstructions, showing that it is no longer circular, but still highly non-local. Furthermore, in Appendix \ref{sec:app2} we dig into the reconstruction problem from the point of view of invertibility of the algorithm, both for CP and ART, and other theoretical aspects.

Regarding non-Cartesian acquisitions, $\kappa$-space is not a meaningful reciprocal to image space (see Appendix\,\ref{sec:app3}), since resolution (and FOV) are not globally defined, as in PatLoc sequences\cite{patloc2008}.

\section{Conclusions}\label{sec:conclusions}

We have introduced SPDS, a magnetic-field mapping method based on double shot SPRITE sequences. Compared to double shot GRE, SPDS does not suffer from incorrect coordinate assignment and it is more robust against eddy currents and SNR drops, lending itself to situations where the main magnetic field is largely imperfect. We have shown that this method provides reasonable field estimations in situations with inhomogeneities as large as 1,000\,ppm over circular samples of only 2\,cm radius (Fig.~\ref{fig:fig_SPDSexamples}d). We have also shown that model-based reconstruction by ART (Kaczmarz's algorithm), with the SPDS field estimate as prior, is able to remove artifacts up to levels approaching the hard reconstructability limit, i.e. when the encoding gradient is no longer able to spectrally resolve the set of Larmor frequencies of spins in two adjacent voxels. We have approached this limit both with linear and non-linear inhomogeneities, and found that almost up to 2,000\,ppm non-linear inhomogeneity in a phantom filling a FoV of (6\,cm)$^2$ can be corrected (Fig.~\ref{fig:fig_simulaciones}e). For linear inhomogeneities, we have seen that one can reconstruct a recognizable image {\it beyond} the reconstructability limit, even if some artifacts are apparent. On the contrary, Conjugate Phase reconstruction and Fourier methods fail.

SPDS field estimation, being based on point-wise encoding, can take longer than a standard double-shot GRE, but it might be the only option for special-purpose hardware. In this case, iterative algorithms can be used to obtain a non-distorted field map \cite{Fessler2004,Koolstra2021}, while SNR drops need to be recovered by expensive increases of signal averaging. In addition, we have shown that low-resolution $\Delta B_0$ field maps usually suffice, and the problems encountered with our method had to do with borders of the phantom (and the masking/unwrapping steps) used to map the field. The method lends itself e.g. to daily calibrations of the field, where an homogeneous sample fills the full region of interest, so that later any other sample or patient is smaller than the map obtained, as long as local permeability effects in the sample can be disregarded.

This work stemmed from the need to reconstruct images from a highly inhomogeneous scanner designed for \emph{in vivo} dental imaging, but its impact can be extrapolated to most LF-MRI systems, where $B_0$ tends to be less homogeneous than in high-field scanners and related artifacts are usual \cite{Koolstra2021}. In Cartesian acquisitions they can be described by geometric distortions (i.e. they are equivalent to coordinate transformations), but in non-Cartesian sequences inhomogeneities lead to blurring, brightness concentration and other deleterious effects. Furthermore, even more disruptive designs, such as gradient-free scanners \cite{Cooley2015} or hand-held devices \cite{GREER2019106591,s17030526}, seem to lead to an increasing level of experimental imperfections that demand advanced reconstruction algorithms \cite{deLeeuw2022}. For all these reasons, it is desirable to find robust methods to cope with strong field inhomogeneities and non-Cartesian sequences.

\section*{Author contributions}
All authors contributed equally to this work.

\section*{Acknowledgment}
We thank present and former members of the MRILab team for their support and participation in the design and assembly of the experimental setup. This work was supported by the Ministerio de Ciencia e Innovaci\'on of Spain through research grant PID2022-142719OB-C22 and the European Innovation Council (EIC-Transition 101136407). Action co-financed by the Agencia Valenciana de la Innovaci\'on (INNVA1/2022/4 and INNVA1/2023/30). JB acknowledges support from the Innodocto program of the Agencia Valenciana de la Innovaci\'on (INNTA3/2021/17).

\appendix

\subsection{Prospective dental 197\,mT scanner}
\label{sec:app}

\begin{figure}[h!]
	\includegraphics[width=0.5\textwidth]{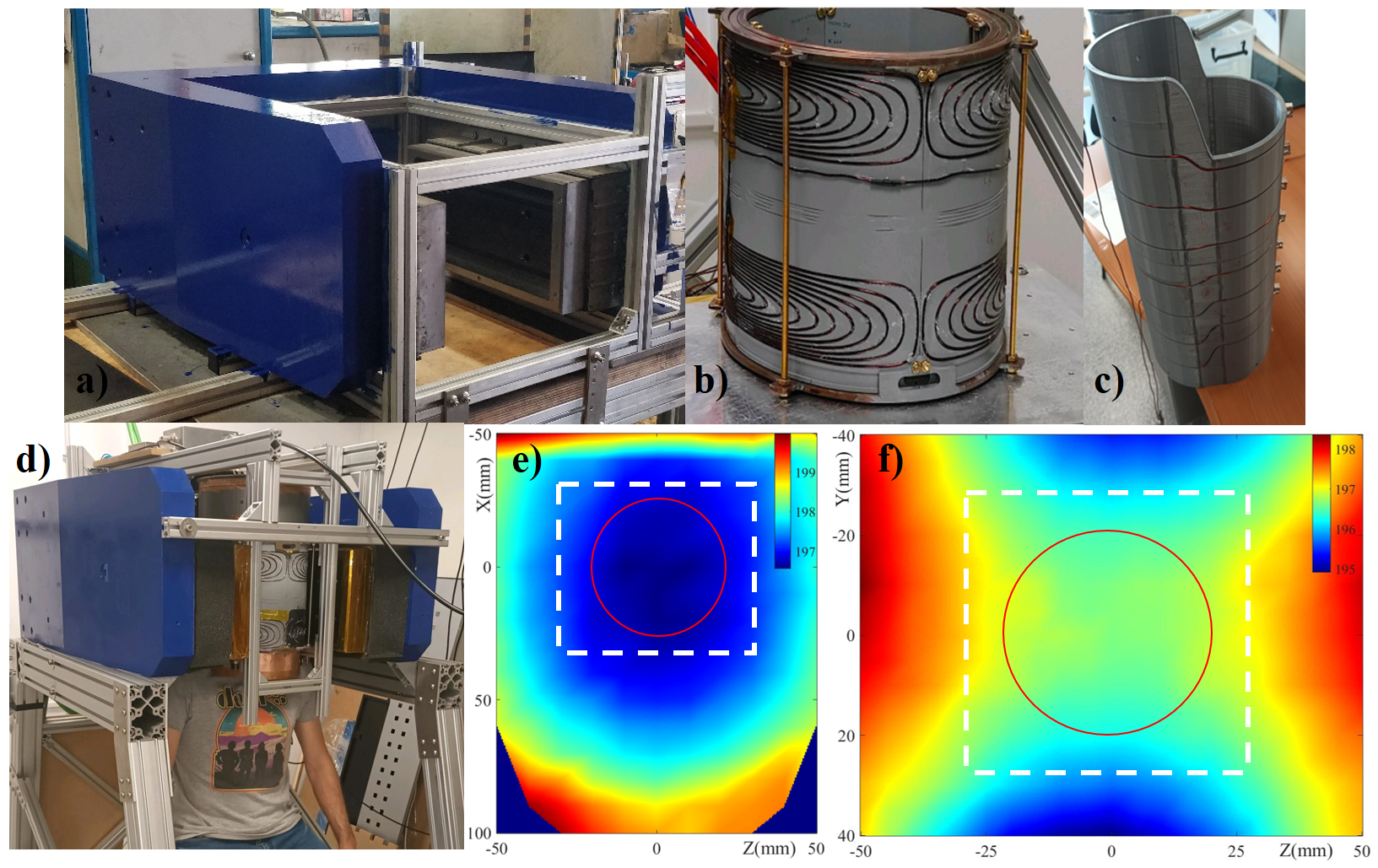}
	\caption{197\,mT scanner designed for low-field dental MRI. a) Yoked magnet with shimming units. b) Water-cooled gradient system. c) RF excitation coil. d) Fully assembled scanner. e) Principal magnetic field $B_0$ in the $xz$ plane ($y=0$). f) Field in the $yz$ plane ($x=0$). The square represents the (6\,cm)$^2$ FoV, while the circle represents the 4\,cm diameter i3M phantom. }
	\label{fig:Dental2scanner}
\end{figure}
Our group has recently demonstrated the viability of \emph{ex vivo} imaging of human teeth in sub-tesla fields \cite{Algarin2020}. To eventually tackle the considerable remaining challenge of \emph{in vivo} dental imaging, we have built a 197\,mT scanner with a C-shaped yoked permanent magnet. Constrained by the human anatomy and restrictions on maximum weight, the magnet poles are small in size and highly asymmetric (see Fig.\,\ref{fig:Dental2scanner}). This leads, even after passive shimming, to a strongly inhomogeneous $B_0$ field. Similar geometric constraints apply to the gradient coils, which are inside the magnet poles for efficiency.

The main field is generated by a home-made magnet built with an array of NdFeB N50-grade cubes with size $42\times42\times95$\,mm$^3$, for a total weight of 1,200\,kg. The gap between poles is 27\,cm and the height is only 24\,cm to place the most homogeneous region inside the patients' mouths. After passive shimming by magnetic cubes on the inner surface of poles, the resulting inhomogeneity is 14,000\,ppm in a region compatible with clinical practice ($92\times71\times110$\,mm$^3$).

The gradient coils have elliptical cross sections to accommodate a human head. They have been optimized with conventional target-field methods, yielding efficiencies of 0.76, 0.27 and 0.51\,mT/m/A, resistances of 102, 248 and \SI{123}{\ohm} and inductances of 111, 94 and \SI{70}{\micro H},  with the idea to drive 40\,A of maximum DC current, due to thermal dissipation. The wires are pressed to a holder made with copper for cooling, where the holder is welded to three conduits with a continuous flow of 4\,l/min supplied by an external water chiller. The gradient coils are driven by a 400\,A and 750\,V gradient power amplifier (International Electrics Co.), even if we limit the maximum current to 100\,A due to the significant duty cycle typical of ZTE sequences.

The RF transmission antenna is a solenoid wound on a 3D printed PLA support, with seven turns and one gap capacitor per loop. Once tuned to 8.36\,MHz and matched to \SI{50}{\ohm}, it has a $Q\approx52$. High power RF pulses are produced by a commercial power amplifier (Barthel HF-Technik GmbH). We typically drive hard $\pi/2$ pulses in \SI{15}{\micro s}, which corresponds to $\approx 60$\,kHz or around 8,000\,ppm, enough to span our complete 3D phantoms.

We employed various single-channel receivers, either solenoids or planar coils of different diameters, depending on the sample.

\subsection{Limits of the reconstruction problem}
\label{sec:app2}

Let us dig a bit deeper into the limits of reconstruction under large inhomogeneities by going to the worst case scenarios of our experiments, that is Fig.\,\ref{fig:fig_SPDSexamples}e) and Fig.\,\ref{fig:fig_ARTvsCP} (right). In both cases, the intravoxel bandwidth reaches the level of encoding bandwidth 625\,Hz/voxel. Reconstructability can be cast as whether the MRI signal samples coming from the experiment can fully describe the sample under study. Each signal sample $s(\vec{\kappa}_i)$ interrogates the sample through the linear transformation given by $\vec{s}=E\,\vec{\rho}$, or equivalently, each measurement $\vec{s}_i=s(\vec{\kappa}_i)$ quantifies the contribution of each `distorted' plane wave $w_i$ $$w_i(\vec{x})=\exp{-\I[\vec{\kappa}_i\,\vec{x}+\gamma\Delta B_0(\vec{x})t(\vec{\kappa}_i)]}$$ to the description of our sample. Thus, the question about reconstructability can be posed as the problem of whether, for a given number of measurements (samples) and resolution (pixels): i) our measurement basis is sufficient for that resolution; and ii) the number of measured samples is enough. While we believe this problem deserves careful and dedicated attention, we will try here to answer, at least partially, the first question, while leaving the second for a more thorough future study.

The spatial basis $\{ \vec{w}_i \}$ (with $(\vec{w}_i)_j=w_i(\vec{x}_j)$) would ideally describe our sample with a given spatial resolution if it is orthogonal, i.e. $\vec{w}_i\cdot\vec{w}_j^*=\mathbb{1}_{i,j}$ with $\mathbb{1}$ the identity, and complete, i.e. $\sum_\kappa\left(\vec{w}_\kappa^H\vec{w}_\kappa\right)_{i,j}= \mathbb{1}_{i,j}$ (with $^H$ the conjugate transpose). The first condition ensures that each vector $\vec{w}_i$ contributes independent, perfectly distinguishable information through their orthogonality. The second condition ensures that the set of vectors $\{ \vec{w}_i \}$ spans completely the space (in this case of sample images) that we are trying to describe, because the sum of projectors is the identity in the target space (image).Typically, when a vector basis does not fulfill orthogonality among its vectors, and the dimension is equal to that of the target space (in MRI, this is commonly called Nyquist or fully sampled), completeness is compromised, but the set of vectors can be usually enlarged to ammend completeness.

Noticing that $w_i(\vec{x}_j)=E_{i,j}$, i.e. the encoding matrix element $(i,j)$, orthogonality can be written as $E\cdot E^H=\mathbb{1}$, and completeness as $E^H\cdot E=\mathbb{1}$. ART reconstruction is known to be equivalent to the $l_2$-norm problem of minimzing $F=||\vec{s}-E\,\vec{\rho}||_2^2$ \cite{randART2013}, whose solution can be shown to be the Moore-Penrose pseudo-inverse matrix $E^+=(E^H\, E)^{-1} \, E^H$, so that $\vec{\rho}=E^+\vec{s}$. Clearly, invertibility, i.e. existence of an image $\vec{\rho}$ compatible with the measured signal $\vec{s}$, is compromised if the eigenvalues of $E^H\, E$ are close to zero (i.e. if the completeness matrix has near null spaces).  

For the rest of the Appendix we use a Cartesian $\kappa$-space (but still with an inhomogeneity term like $|\vec{\kappa}|\Delta B_0(\vec{x})/G$) for simplicity, and acquisitions correspond to 100$\times$100 pixels in a FoV of 6\,cm$^2$. In the linear case we assume the field model $\Delta B_0(z)=0.03\,z$ (i.e. 30\,mT/m), and for the case of curvatures $\Delta B_0(y,z)=-0.7\,y^2+0.5\,z^2$.

\begin{figure}
\begin{center}
	\includegraphics[width=0.4\textwidth]{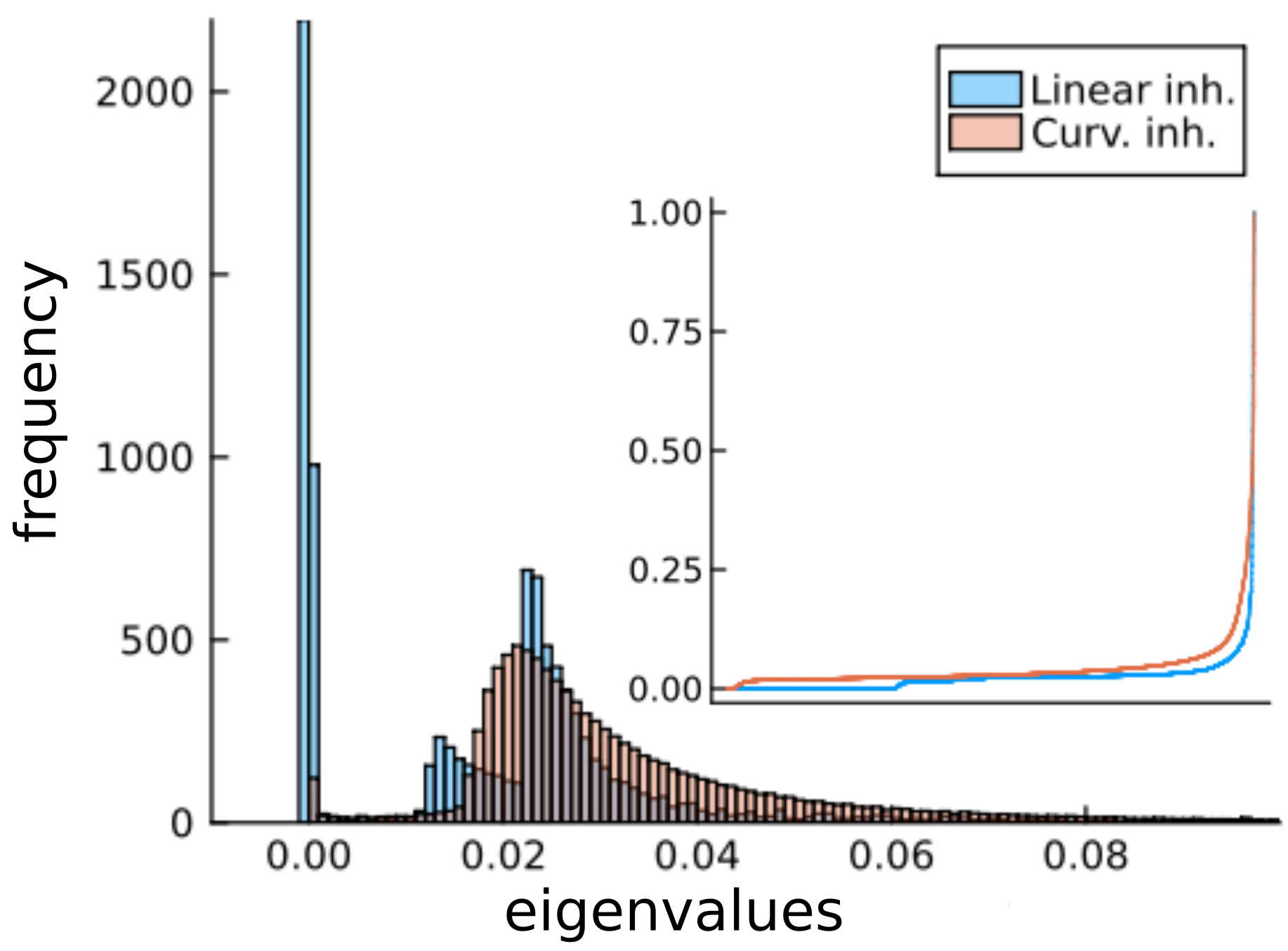}
\end{center}
	\caption{Histogram of eigenvalues of matrix $E^H\,E$ in the Cartesian case for curvatures and linear inhomogeneities, with $100\times100$ pixels. Inset: The same eigenvalues sorted.}
	\label{fig:fig_ART_eigen}
\end{figure}

In Fig. \ref{fig:fig_ART_eigen} we plot the eigenvalues of the matrix $E^H\, E$, normalize eigenvalues to their maximum value, for the cases of Fig.\,\ref{fig:fig_SPDSexamples}e and Fig.\,\ref{fig:fig_ARTvsCP} (right), i.e. with curvature and linear inhomogeneities respectively. It can be seen that a great amount of eigenvalues drop near, or even are, zero when full inhomogeneity is present (at full FoV). In particular, for linear inhomogeneity above the encoding gradient, many eigenvalues drop to zero.

\begin{figure}[h!]
	\begin{center}
	\includegraphics[width=0.4\textwidth]{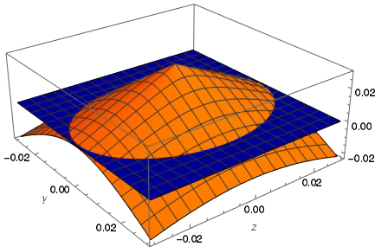}
	\end{center}
	\caption{Jacobian of the transformation $\{\vec{\kappa}\}\rightarrow\{t,\phi\}$, for case of curvatures (orange), with the 0 value in blue for comparison.}
	\label{fig:fig_Jacobian_YZ}
\end{figure}

Unlike the case of linear inhomogeneity, field curvatures affects most prominently parts of the sample in the outer regions of the FoV, but this information is not visible in the eigenvalues of $E^H\, E$. However, the authors of \cite{CP2005} proposed a condition to check invertibility of the Conjugate Phase reconstruction method, by analyzing the Jacobian of the transformation $\{\kappa_x,\kappa_y\}\rightarrow \{t,\phi\}$. This Jacobian needs to be positive for CP to be invertible (without a change of orientation in the transformation). The Jacobian for the linear inhomogeneity case is $$D^\text{linear}(t,\phi)=G\,t(G+g_\text{inh.}\cos\phi)$$ and for the case with curvatures $$D^\text{curv.}(t,\phi,y,z)=G\,t(G+2\alpha y\cos\phi+2\beta z\sin\phi)$$ with $(\alpha,\beta)$ the curvatures in $(y,z)$ respectively. In the linear case, invertibility is compromised when $G\lesssim g_\text{inh.}$ (and $\phi=\pi$). In the case of curvatures, we plot in Fig.\,\ref{fig:fig_Jacobian_YZ} the minimum value of the Jacobian with respect to $\phi$ at each point $(y,z)$, that is $$\min_{\{\vec{\kappa}\}\equiv \{(t,\phi)\}}D^\text{curv.}(t,\phi,y,z).$$ In this case, the invertibility depends on the pixel, thus giving a maximum region where invertibility is expected to hold.

An interesting extension would be to find a similar criterion for invertibility of the ART algorithm. A reasonable starting point would be to analyze the properties of the pseudo-inverse matrix elements $(E^+)_{\vec{x}_0,\vec{\kappa}}$ evaluated at a given FoV point $\vec{x}_0$, but this is left for future exploration.

\begin{figure}[h!]
	\begin{center}
	\includegraphics[width=0.3\textwidth]{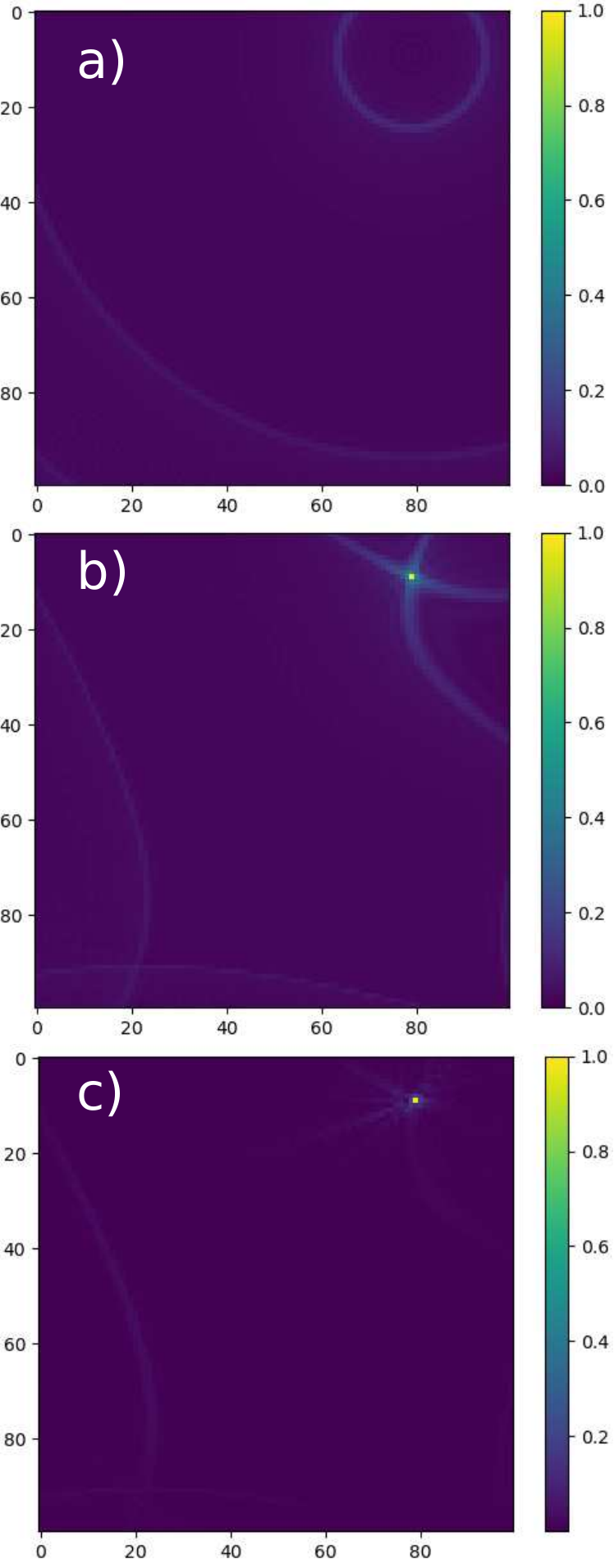}
	\end{center}
	\caption{Point spread function in the curvatures case, for CP without (top) and with PK (middle), and for ART reconstruction (bottom). }
	\label{fig:fig_PSF}
\end{figure}

The power of prior knowledge for reconstruction can be seen also in the point spread function (PSF), which is known to cause circular patterns when no prior is applied \cite{NollReview2022}. When the field prior is used, the PSF improves, but still has some non-local artifacts, as seen in Fig.\,\ref{fig:fig_PSF}. There we show the CP reconstruction of a point-like sample with and without PK in the curvatures case, and at the bottom the ART reconstruction with PK for the same sample. CP indeed localizes the PSF much better than Fourier or non-PK reconstruction, but ART clearly outperforms both, while still having non-local image residuals.

Finally, we have checked the improved model for reconstruction as given in \cite{intravoxel2020} where the encoding matrix contains the integration of the inhomogeneity exponent within each pixel. To do so, the inhomogeneity is expanded to first order at the center of the voxel, so that the encoding matrix element for that pixel $(i,j)$ in e.g. the case of curvatures is
\begin{equation}
\int_{-dx/2}^{dx/2}dx \int_{-dy/2}^{dy/2}dy e^{-\I\vec{\kappa} \vec{x}_i -\I\frac{|\vec{\kappa}|}{G}(\Delta B_0(x_i,y_i)+2\alpha x_i x+2\beta y_i y)}
\end{equation}
Thus, the encoding matrix for that pixel is
\begin{equation}
E(\vec{\kappa},\vec{x}_i)=\text{sinc}(\frac{\tilde{\kappa}_x dx}{2})\text{sinc}(\frac{\tilde{\kappa}_y dy}{2})e^{-\I\vec{\kappa}\cdot\vec{x}_i -\I\frac{|\vec{\kappa}|}{G}\Delta B_0(\vec{x}_i)}
\end{equation}
and $$\tilde{\kappa}_x=\kappa_x+\frac{2|\vec{\kappa}|}{G}\alpha x_i$$ $$\tilde{\kappa}_y=\kappa_y+\frac{2|\vec{\kappa}|}{G}\beta y_i,$$ for curvatures $(\alpha,\beta)$ along $(x,y)$ directions.

\begin{figure}[h!]
	\begin{center}
	\includegraphics[width=0.2473\textwidth]{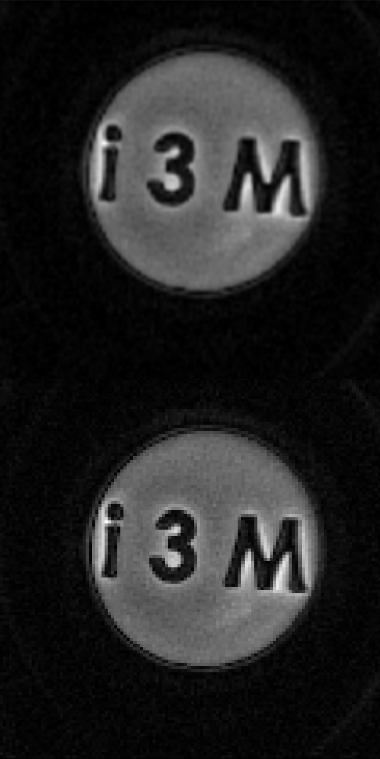}\includegraphics[width=0.25\textwidth]{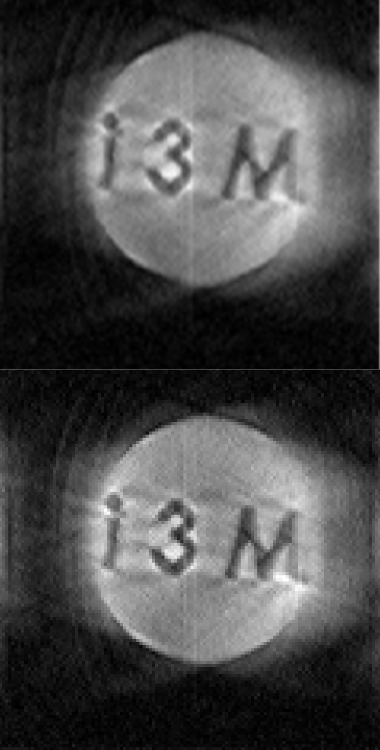}
	\end{center}
	\caption{(Left) Top/bottom: ART reconstruction without/with intravoxel modelling int the case of BW$_\text{voxel}=106$\,Hz linear inhomogenity. (Right) Top/bottom: ART reconstruction without/with intravoxel modelling int the case of BW$_\text{voxel}=639$\,Hz linear inhomogenity.}
	\label{fig:fig_SINC}
\end{figure}

Using the equivalent expression for linear inhomogeneity $$\tilde{\kappa}_x=\kappa_x+\frac{|\vec{\kappa}|g_\text{inh.}}{G}$$ $$\tilde{\kappa}_y=\kappa_y$$ we see a slight improvement for all levels of inhomogeneity. In Fig.\,\ref{fig:fig_SINC} we show the case with BW$_\text{voxel}=106$\,Hz of Fig.\,\ref{fig:fig_ARTvsCP}, and see that edges are better defined.

However, for the highest inhomogeneity ($g_\text{inh.}>G$, BW$_\text{voxel}=639$\,Hz), although edges are better defined, that is of no use to recover an artifact-free image (see Fig.\,\ref{fig:fig_SINC} right)
\begin{figure}[h!]
	\begin{center}
	\includegraphics[width=0.3\textwidth]{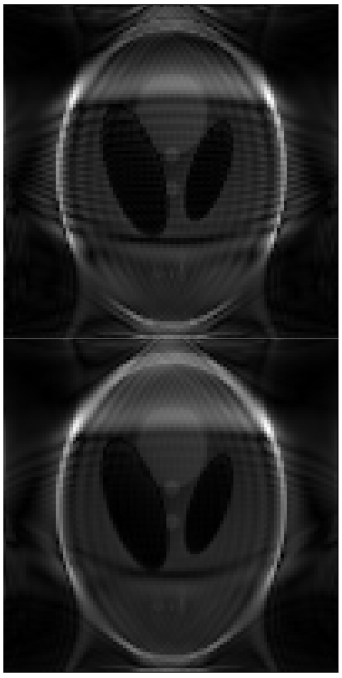}
	\end{center}
	\caption{Top/bottom: ART reconstruction without/with intravoxel modelling in the case of curvatures inhomogenity in the $yz$ plane, for a Shepp-Logan phantom with 100 spins per voxel.}
	\label{fig:fig_SINC_curv}
\end{figure}

For the quadratic field, improvements are seen for a simulated Shepp-Logan (100 spins per pixel) in Fig.\,\ref{fig:fig_SINC_curv} for the case of 197\,mT scanner in the plane $yz$ ($\text{curvatures}=[-0.7,0.5]$), but not significantly for the case of our experimental measurements. This is probably because we did not consider a sinc function along the third dimension. In any case, the strong artifacts cannot be ammended.

\begin{figure}[h!]
	\begin{center}
	\includegraphics[width=0.32\textwidth]{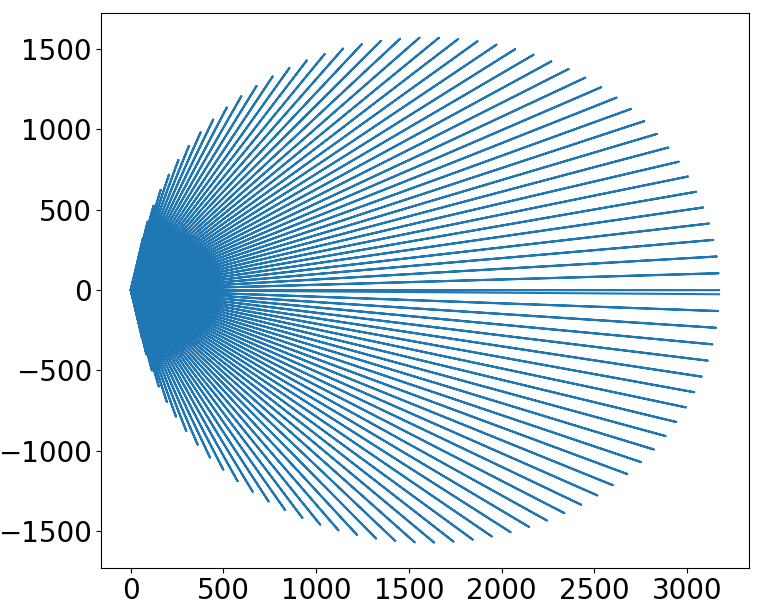}
	\end{center}
	\caption{Local $\kappa$-space for the linear inhomogeneity case, which is independent of image space position.}
	\label{fig:fig_local1}
\end{figure}

\begin{figure}[h!]
	\begin{center}
	\includegraphics[width=0.4\textwidth]{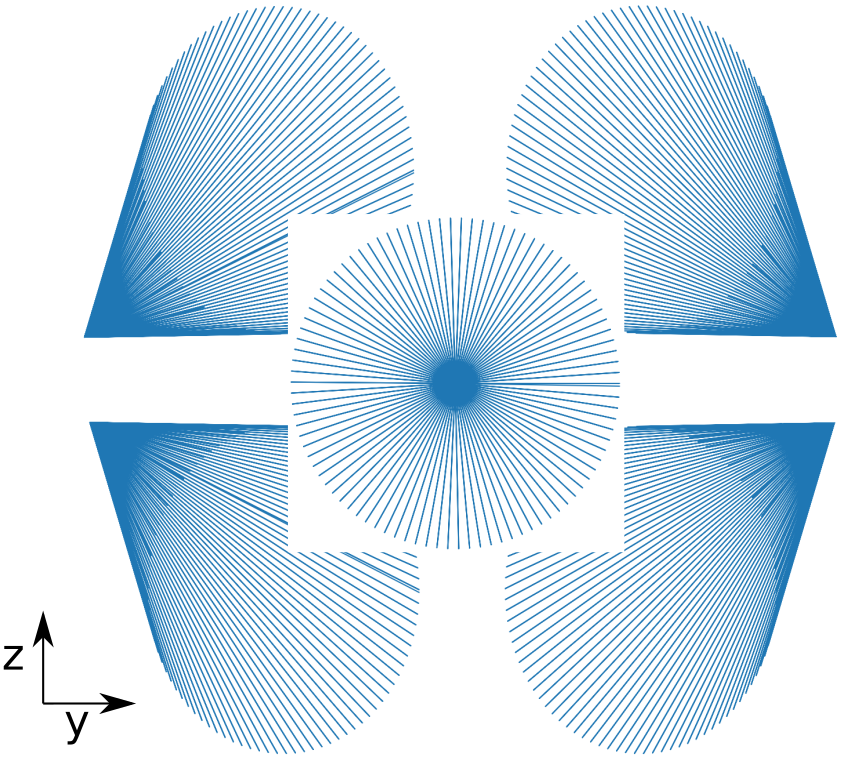}
	\end{center}
	\caption{Local $\kappa$-space for the curvatures (-0.7,0.5) inhomogeneity case, for the corners of image space ($y=\pm 3$cm, $z=\pm 3$cm) and its center. }
	\label{fig:fig_local2}
\end{figure}

As a final note, an interesting concept is that of `local $\kappa$-space' \cite{localKspace2011}, where each point in sample/image space suffers a different $\kappa$-space, through the relation:
\begin{equation}
 \vec{\kappa}_\text{local}(\vec{x},t)=\vec{\nabla}\phi(\vec{x},t),
\end{equation}
where $\phi$ is the acquired phase at each point (note that in our case the expressions $\vec{\kappa}_\text{local}=(\tilde{\kappa}_x,\tilde{\kappa}_y)$, i.e. coincide with those given above). We plot in Figs.\,\ref{fig:fig_local1} and \ref{fig:fig_local2} the local $\kappa$-space with linear and quadratic inhomogeneities, as done in Ref.\,\cite{localKspace2011}. In both cases (for sample points away from the center), we see that the center of local k-space is very partially filled. This indicates that even the grosser spatial details of the image are necessarily destroyed.


\subsection{$\kappa$-space as a bad reciprocal of image space}\label{sec:app3}
Consider a perfectly homogeneous $B_0$ field, where we can return to the use of $k$ instead of $\kappa$. In this case, ZTE provides a description of the sample by a combination of radially-oriented plane waves, with maximum spatial resolution given by the highest wavenumber $k_\text{max}=\gamma g_\text{enc}T_\text{acq}$. For each spoke $i$, the wave is $\exp[-\I(k_\text{max}\cos\theta_i x+k_\text{max}\sin\theta_i y)]$, where the wave is oriented at angle $\theta_i$. When a linear gradient is present, the description of the sample, based on the prior e.g. $\Delta B_0=g_\text{inh.}\cdot x$, waves are distorted: $$\exp[-\I\,k_\text{max}(\frac{g_\text{inh.}}{G_\text{readout}}+\cos\theta_i) x-\I\,k_\text{max}\sin\theta_i\, y)].$$ Such wave has a wave vector modulus which is no longer $k_\text{max}$, but effectively $$k_\text{max}^\text{eff.}= \sqrt{1+\alpha^2+2\alpha\cos\theta_i}\,\,\, ,\,\,\, \alpha=\frac{g_\text{inh.}}{G_\text{readout}},$$ i.e. its modulus (the inverse spatial resolution) depends on the spoke ($k_\text{max}^\text{eff.}=k_\text{max}^\text{eff.}(\theta_i)$). Furthermore, for the extreme case $g_\text{inh.}=G_\text{readout}$ a simple plot reveals that roughly 1/3 of the spokes has $k_\text{max}^\text{eff.}<k_\text{max}$, i.e. lower resolution than expected, 16\,\% of the spokes have half the expected resolution, 8\,\% have a quarter of the expected resolution, etc. reaching $k_\text{max}^\text{eff.}=0$ for $\theta_i=\pi$. The same argument is valid for $dk^\text{eff.}=k_\text{max}^\text{eff.}/N=2\pi/$FOV and therefore, the FOV can be seen as spoke-dependent.

\bibliography{myrefs}
\end{document}